\documentclass{vgtc}                          % final (conference style)
\usepackage{pifont}

\newcommand{\head}[1]{\vspace{0.5mm}\noindent{\bf #1:}}

\usepackage{xspace}

\newcommand{\circled}[1]{\raisebox{.5pt}{\textcircled{\raisebox{-.9pt} {#1}}}}

\ifpdf%                                % if we use pdflatex
  \pdfoutput=1\relax                   % create PDFs from pdfLaTeX
  \usepackage{graphicx}                % allow us to embed graphics files
  \DeclareGraphicsExtensions{.pdf,.png,.jpg,.jpeg} % for pdflatex we expect .pdf, .png, or .jpg files
\else%                                 % else we use pure latex
  \usepackage{graphicx}                % allow us to embed graphics files
  \DeclareGraphicsExtensions{.eps}     % for pure latex we expect eps files
\fi%

\graphicspath{{figures/}{pictures/}{images/}{./}} % where to search for the images

\usepackage{subcaption}
\usepackage{microtype}   
\PassOptionsToPackage{warn}{textcomp}  % to address font issues with \textrightarrow
\usepackage{textcomp}                  % use better special symbols
\usepackage{mathptmx}                  % use matching math font
\usepackage{times}                     % we use Times as the main font
\usepackage{cite}                      % needed to automatically sort the references
\usepackage{tabu}                      % only used for the table example
\usepackage{booktabs}                  % only used for the table example
\usepackage{array}
\newcolumntype{C}[1]{>{\centering\arraybackslash}p{#1}}
\usepackage{svg}
\usepackage{amsmath}

\title{LiveVV: Human-Centered Live Volumetric Video Streaming System}

\teaser{
  \centering
  \includegraphics[width=\linewidth,trim=45 0 45 0]{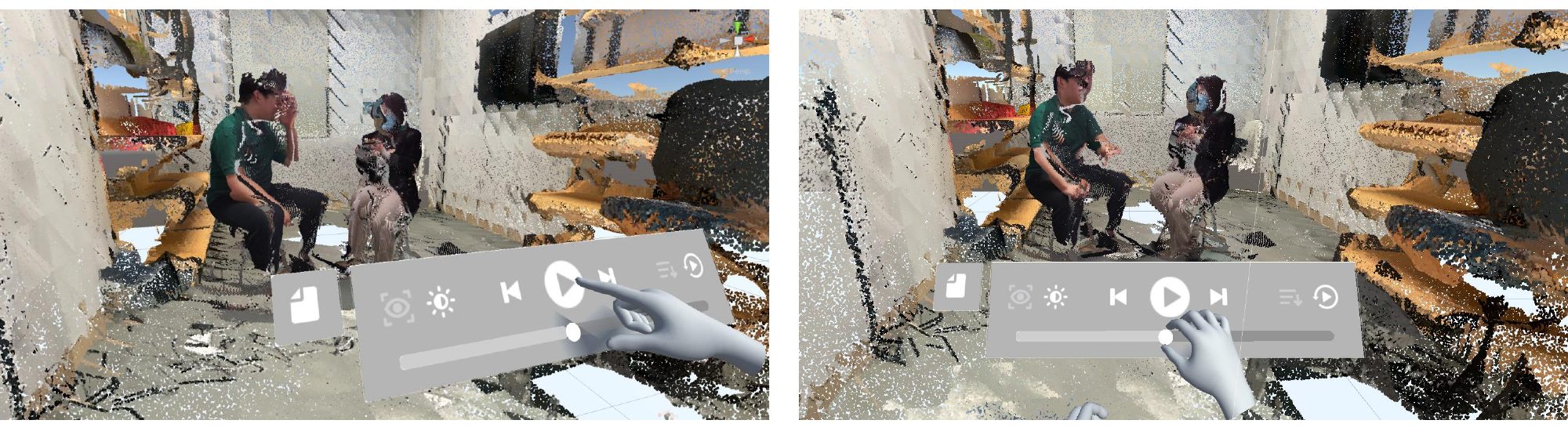}
    \vspace{-0.7cm}  
  \caption{User's view when using LiveVV to watch a live streaming volumetric video}
  \label{fig:teaser}
}

\abstract{
Volumetric video has emerged as a prominent medium within the realm of eXtended Reality (XR) with the advancements in computer graphics and depth capture hardware. Users can fully immersive themselves in volumetric video with the ability to switch their viewport in six degree-of-freedom (DOF), including three rotational dimensions (yaw, pitch, roll) and three translational dimensions (X, Y, Z). Different from traditional 2D videos that are composed of pixel matrices, volumetric videos employ point clouds, meshes, or voxels to represent a volumetric scene, resulting in significantly larger data sizes. While previous works have successfully achieved volumetric video streaming in video-on-demand scenarios, the live streaming of volumetric video remains an unresolved challenge due to the limited network bandwidth and stringent latency constraints.

In this paper, we for the first time propose a holistic live volumetric video streaming system, LiveVV, which achieves multi-view capture, scene segmentation \& reuse, adaptive transmission, and rendering. LiveVV contains multiple lightweight volumetric video capture modules that are capable of being deployed without prior preparation. To reduce bandwidth consumption, LiveVV processes static and dynamic volumetric content separately by reusing static data with low disparity and decimating data with low visual saliency. Besides, to deal with network fluctuation, LiveVV integrates a volumetric video adaptive bitrate streaming algorithm (VABR) to enable fluent playback with the maximum quality of experience. Extensive real-world experiment shows that LiveVV can achieve live volumetric video streaming at a frame rate of 24 fps with a latency of less than 350ms. 

} % end of abstract

\begin{document}

%% The ``\maketitle'' command must be the first command after the
%% ``\begin{document}'' command. It prepares and prints the title block.

%% the only exception to this rule is the \firstsection command
%\firstsection{Introduction}

\maketitle

\begin{figure*}[ht]
    \begin{minipage}[t]{0.50\textwidth}
    \end{minipage}
\end{figure*}

%% \section{Introduction} %for journal use above \firstsection{..} instead
\section{Introduction}
%2.0-pages with abstract
%//TODO: callback Vivo and Cav3.
%paragraph 1: background of VV
Volumetric video (VV) emerges as one of the most representative immersive media paradigms with the advances in computer graphics and depth capture hardware, allowing users to fully immerse themselves in a virtual 3D video scene and freely switch their viewport by wearing a head-mounted display. Compared with traditional 2D videos and 360 videos, volumetric video provides users with six degree-of-freedom (DOF), besides three rotational dimensions (yaw, pitch, roll) of 360 videos, it further provides three translational dimensions (X, Y, Z). With such an ability to provide immersive experiences, volumetric video is expected to become a killer application in the new era of `spatial computing', and will empower services in VR, AR, and MR. It is estimated that the global volumetric video market will reach 7.6 billion USD by 2028~\cite{VV_market}.

%paragraph 2: data format description
Unlike traditional 2D or 360 videos that are composed of pixel matrices, volumetric videos are represented by multiple formats: textured mesh~\cite{VV_format_mesh}, voxel~\cite{VV_format_voxel}, V-PCC~\cite{VV_format_VPCC}, or colored point cloud~\cite{VV_format_PtCl}. Due to the feature of simplicity and easy manipulation, point cloud is the most popular data format. Despite the advantage of immersive experience and high interactivity, the enormous data size and limited network bandwidth capacity hinder the delivery of volumetric videos in application scenarios, not to mention live streaming. Taking a full-scene volumetric video as an example, when the frame rate is 30 fps and the total number of points per frame is about 920,000, the bandwidth requirement is up to 11.1 Gbps~\cite{FSVVD}.

%paragraph 3: related previous work
Currently, existing works about volumetric videos mainly focus on the optimization of streaming pre-recorded volumetric video content, i.e. video-on-demand. Major solutions include tile-based adaptive volumetric video streaming, super-resolution-based volumetric video streaming, and layered streaming. For example, ViVo~\cite{ViVo} encoded tiles into different qualities according to their relationship with users’ viewport, distance, and occlusion and designed a simple adaptive streaming scheme. While YuZu~\cite{YuZu} exploited 3D super-resolution models to increase visual quality in volumetric video streaming. However, none have achieved a holistic system that is able to capture, process, and deliver volumetric video content in real-time. Such a holistic system holds the potential for more captivating application scenarios than its on-demand counterpart, like telepresence, remote education, or immersive entertainment.

%paragraph 4: describe our work 
%challenges
Through investigating previous related works, we conclude the challenges for building a holistic live volumetric video streaming system in three aspects:

First, current volumetric video capture methods~\cite{FSVVD, CWIPC_SXR} are heavy and complicated. Normally volumetric videos are captured by an array of depth cameras from multiple angles, thus necessitating meticulous preparatory work during deployment to calibrate and synchronize the cameras in order to get a holographic view. For instance, FSVVD~\cite{FSVVD} entails positioning a reference structure at the center of the target region prior to capture for calibration. However, such an approach is impractical for live-streaming scenarios where capture devices must be quickly and easily deployed.

%First, currently live volumetric video capture has not been achieved. Normally, volumetric videos are captured by an array of depth cameras from multiple angles, thus it's necessary to calibrate and synchronize the cameras in order to get a holographic view. However, for live streaming scenarios, we need to perform the above process in real-time, which poses challenges to algorithm optimization and computing power.

Second, a huge gap exists between the limited network transmission bandwidth and the enormous volumetric video data size. Volumetric videos are normally represented by textured meshes or colored point clouds, that are a lot more larger in data size compared with traditional video. Previous works on video-on-demand scenarios mainly focus on optimizing the streaming of pre-recorded volumetric videos, in which optimizations like compression and buffering can be exploited. For live streaming scenarios, volumetric video data must be processed in real-time, thus making previous methods not applicable. %we need to figure out new methods to control the bandwidth consumption.

%Third, maintain the best possible visual quality in fluctuating network conditions
Third, fluctuating network condition poses a great challenge for live video streaming tasks, especially for volumetric videos that have a huge bandwidth requirement. Different from traditional flat videos, simply adapting the resolution according to variating network conditions is not feasible for volumetric video streaming, since such reckless processing has a large impact on the perceived visual quality. As examples shown in Figure~\ref{fig: different_PDL}, we can observe a huge visual quality drop after global decimation.

%Paragraph 5: We propose  LiveVV to address the above concern.
To tackle the aforementioned challenges, we present LiveVV, the pioneering comprehensive live volumetric video streaming system that seamlessly integrates capture, scene reuse, streaming, and rendering. This system comprises four key modules: real-time volumetric video generation across multiple capture devices, scene segmentation \& reuse, volumetric video adaptive bitrate transmission, and client-side volumetric video MR display. These modules collectively form the backbone of LiveVV, enabling the entire pipeline of live streaming volumetric video, from capture to rendering on the client side.

%introduce the system 
\noindent\textbf{Real-time VV Generation} is accomplished through the seamless integration of capture devices and edge processing units. In order to enable the instantaneous creation of volumetric videos and facilitate efficient deployment, we leverage body tracking data to achieve dynamic calibration. Since most depth capture devices are integrated with body tracking functions, we can easily access the body tracking data to perform the calibration. The captured volumetric video content from each angle is calibrated in real-time based on the calibration results.

%\noindent\textbf{Real-time VV Generation} is achieved by the integration of an edge computing unit and an Azure Kinect depth camera, and multiple capture units compose a camera array to capture the RGB and depth data from different angles around the target region. The capture units are connected via a network switch for temporal synchronization to get a synchronized frame. In order to satisfy the rapid lightweight deployment requirement for dynamic scenarios, we exploit the body track skeleton data to calibrate each capture unit dynamically. The capture result is colored point clouds and body track skeleton joints, which will be used for further optimization processing.%后面怎么写再想想（output colored PtCl and skeleton joints）

%to generate colored point clouds and extract body tracking feature joints for further optimization processing. 

\noindent\textbf{Scene Segmentation \& Reuse} segment the data from the capturer array, based on body skeleton tracking data, it performs precise segmentation and decimation to the whole scene according to visual saliency. For each frame, the whole scene is segmented into two parts: dynamic human body and static external scene. The dynamic human body is further segmented into several parts and decimated according to the saliency of each part in order to reduce data size and maintain a better visual quality as well. The static external scene is updated periodically at different frequencies according to the visual saliency and network conditions. 

\noindent\textbf{Bandwidth Adaptation for VV}
is a specially designed bandwidth adaptation scheme for volumetric video streaming in order to deal with network fluctuation. Unlike traditional 2D video streaming optimization, simply switching the resolution according to the bandwidth changes will result in a huge impact on the overall visual quality. Thus we propose a bitrate adaptation mechanism specifically for volumetric video. Considering the inherent characteristics of volumetric videos, we assign distinct levels of point cloud density to various regions within the dynamic components of the volumetric scene. Additionally, we establish varying update frequencies for the static scenes based on network conditions to fully reuse the static scene data with high repetition. This approach aims to sustain a consistent frame rate while ensuring optimal visual quality.

%We first investigate the VV content features and user perception preferences with in-depth analysis to
%paragraph 6: Summarize the contributions
We implement the aforementioned modules using commodity devices and integrate them into a holistic prototype system. This system allows us to evaluate the performance of the modules in real-life live-streaming scenarios. Based on the experimental results, we summarize our contributions as follows:

\noindent$\bullet$ LiveVV achieves live volumetric video streaming at a stable frame rate of 24fps and a latency of less than 350ms, which outperforms the SOTA, LiveScan3D~\cite{kowalski_livescan3d_2015}.

\noindent$\bullet$ LiveVV employs a body-tracking-based dynamic calibration method to achieve a real-time volumetric video frame calibration.

\noindent$\bullet$ By reusing static volumetric content with high repetition, LiveVV significantly reduces bandwidth consumption in live volumetric video streaming.

\noindent$\bullet$ We propose a volumetric video adaptive bitrate streaming algorithm, that maintains the quality of experience under fluctuating network conditions.

\noindent\textbf{Ethical:} This work raises no ethical and anonymity problem.

\section{Background}
% In this section, we first provide a background of an end-to-end volumetric video streaming system, which involves two steps: VV creation and VV streaming. Then, we explain the need to split the dynamic and static content of the video. 
%0.5pages
\subsection{2D Telepresence}
In recent years, there has been a widespread adoption of 2D telepresence systems such as Zoom, Facetime, and MS Teams. However, most of these traditional systems only offer a single 2D view captured from one or more angles. Such approaches create a significant gap between users, limiting immersion and interactivity compared to real-world face-to-face conversations where individuals can freely change their viewpoints.

To address this limitation, various efforts have been made to bridge the gap and enhance the user experience in 2D telepresence system. Notable solutions include optimizing camera placement to synchronize eye contact~\cite{10.1145/503376.503386, 10.1145/1460563.1460593}, situated displays, and avatars~\cite{10.1145/1531326.1531370, 10.1145/2766974}. More advancements also include gaze-preserving multiview telepresence~\cite{10.1145/2556288.2557320} and improvements in gaze estimation for telepresence~\cite{10.1016/j.ijhcs.2015.10.004}.

\subsection{Volumetric Video Creation}
Unlike traditional 2D video that is composed of a pixel matrix, and can be directly acquired by RGB cameras from fixed viewpoints, volumetric video captures a dynamic representation of a scene from multiple angles, providing viewers the ability to explore the scene freely. The creation of volumetric videos involves capturing dynamic scenes from multiple angles to achieve three-dimensional reconstruction, which requires an array of depth cameras positioned around the target capture region. The captured information, RGB images, and depth maps are typically processed to generate an initial representation of 3D scenes from different angles. After multi-camera calibration and temporal synchronization, the scenes from different angles could be aligned to construct a complete volumetric video. Most popular volumetric video representation formats include colored point cloud~\cite{DBLP:journals/pami/GuoWHLLB21}, textured mesh~\cite{DBLP:conf/imr/Owen98,DBLP:journals/cgf/BommesLPPSTZ13}, and voxel~\cite{xu2021voxel}.

\subsection{Volumetric Video Streaming}

% \textbf{What are full-scene VV live streaming and its motivating scenarios}

One-to-many live streaming is useful in e-commerce, telepresence, and remote MOOC education. To maximize the user viewing experience, it requires streams of both the background scene and the human body. For example, in online teaching, both the teacher and the blackboard must be streamed to the user. It also requires a simple and robust setup for the stream host.

% Recent research has explored various methods for volumetric video streaming: Cav3~\cite{Cav3} leverages buffering and field-of-view (FoV) prediction to optimize the streaming process, ViVo~\cite{ViVo} uses a FoV-based approach to minimize the amount of points to be transmitted to the audience, but the system did not consider the real-time capture scenario. YuZu~\cite{YuZu} utilizes a neural-based approach to compress the point clouds for streaming, but it requires pre-training the model for each video, which costs hours of training time. Holoportation~\cite{Holoportation} uses a lightweight compression to compress the texture and geometry to provide real-time scene capture, but it requires 1-2 Gbps bandwidth for each scene.

Recent research has explored various methods for volumetric video streaming. CaV3~\cite{Cav3} leverages buffering and field-of-view (FoV) prediction to optimize the streaming process. ViVo~\cite{ViVo} employs a FoV-based approach to minimize the number of points transmitted to viewers but does not account for real-time capture scenarios. YuZu~\cite{YuZu} uses a neural network to compress point clouds for streaming, however, it necessitates pre-training a model for each video, requiring hours of training. Holoportation~\cite{Holoportation} utilizes a lightweight compression to compress texture and geometry data to enable real-time scene capture, though it demands 1-2 Gbps bandwidth per scene. FarfetchFusion~\cite{lee_farfetchfusion_2023} uses a temporality similarity-based method to stream the 3D face, but due to the limitation of their facial landmark detection mechanism, it is naturally only applicable to the scenario of face-to-face telepresence. While these works have advanced volumetric video streaming, challenges remain in balancing compression, streaming bandwidth, pre-processing time, and real-time performance. Further research can build on these efforts to develop streaming solutions tailored for live capture and interactive viewing across diverse network conditions.
% \begin{itemize}
%     \item One-to-many VV streaming could be used in e-commerce, online MOOC education, etc. The scene and the body are equally important.
%     \item One to many requires a quick setup (a lightweight streaming system.
% \end{itemize}
% \textbf{Problems for VV and full-scene VV live streaming of existing solutions:}
% \begin{itemize}
    % \item High bandwidth consumption: ViVo, X promoted a method X, but it does not fit in the one-to-many live-streaming scenario, because they did not consider the scene, while most of the parts of the scene stay structurally unchanged (measured in S3)
    % \item High processing (computation) time and unable to real-time streaming: YuZu, X, does not fit into this scenario, because they require an expensive setup in streaming (train a model, etc.) which could not be tolerated.
    % \item FSVVD~\ref{}
% \end{itemize}

%FSVVD~\cite{FSVVD} utilizes a physical box calibration technique and requires heavy setup work, which does not fit in the proposed scenario. Project Starline~\cite{lawrence_starline_2021} focuses on the telepresence of the upper half body and also requires specialized capture hardware. LiveScan3D~\cite{kowalski_livescan3d_2015} uses markers to calibrate multiple Kinect v2 sensors and capture in near real-time, but any movement to the sensor will lead to re-setup which is not robust.

% \todo{Categorize: , , }

To the best of our knowledge, none of the existing works have achieved live streaming for full-scene volumetric video due to large bandwidth consumption and high computation costs.

\iffalse
\begin{figure}[t]
    \centering
    \includegraphics[width=\columnwidth]{figures/bandwidth/different_pdl.pdf}
    % \vspace{-0.5cm}
    \caption{Comparison of different PDL}
    \label{fig: different_PDL}
    % \vspace{-0.5cm}
\end{figure}
\fi
\iffalse
 \begin{table}[t]
    \centering
    \begin{tabular}{|c|c|c|c|}
    \hline
        \textbf{Video} & \textbf{Dynamic Human Body} & \textbf{Static Scene} & \textbf{Full Scene} \\ \hline
        \textbf{Video1} & 0.76 & 10.12 & 10.88 \\ \hline
        \textbf{Video2} & 1.50 & 17.36 & 18.86 \\ \hline
        \textbf{Video3} & 0.94 & 15.36 & 16.30 \\ \hline
        \textbf{Video4} & 0.92 & 9.18 & 11.10 \\ \hline
    \end{tabular}
\end{table}
\fi

\section{Measurement and Motivation}
In this section, we conduct in-depth measurements into volumetric video streaming to quantitatively show the challenges, and motivate our human-centered streaming approach.
\label{sec: measurement}

% \begin{itemize}
%     \item We measure how the scene changes in one-to-many live streaming and observe that the body takes the most part changes.
%     \item We then split the body and scene, measured their contribution to the bandwidth consumption, and found that the scene contributes little changes but most bandwidth
%     \item Since the body is the main contributor of the bw now, we further consider is it possible to decrease the bw consumption of the body? Intuitively, complex body structures like head and hand have a greater quality decrease. We use downsample technique for each part of the body and test their visual quality by SSIM.  We measured 1 body-part-based downsampled frame and showed that different body part has different potential for saving bandwidth at a given quality target.
% \end{itemize}

% We make measurements on several full-scene volumetric video content to address the following questions: \textit{(1) How does the dynamic and static content contribute to the bandwidth consumption?} We split the body and scene apart and show that the static content contributes to more than \todo{X} of total bandwidth. \textit{(2) How much does scene change} We validate \todo{} and results gives that \todo{}. \textit{(3) How much opportunity do we have to have each body part downsampled?} We split the body into 5 parts, quantitatively measure their downsampled visual quality to the original one, and show that \todo{}.

\begin{figure*}[t]
    \begin{minipage}[t]{0.3\textwidth}
        \centering
        \includegraphics[width=\columnwidth]{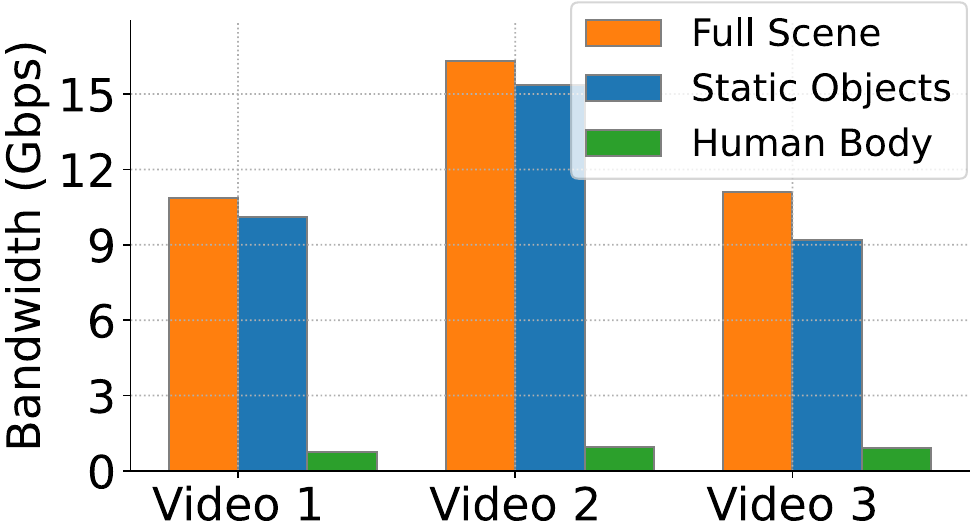}
        % \vspace{-0.5cm}
        \caption{Bandwidth Consumption of streaming volumetric video}
        \label{fig: bandwidth_consumption}
        % \vspace{-0.5cm}
    \end{minipage}
     \hfill
    \begin{minipage}[t]{0.34\textwidth}
        \centering
        \includegraphics[width=0.9\columnwidth]{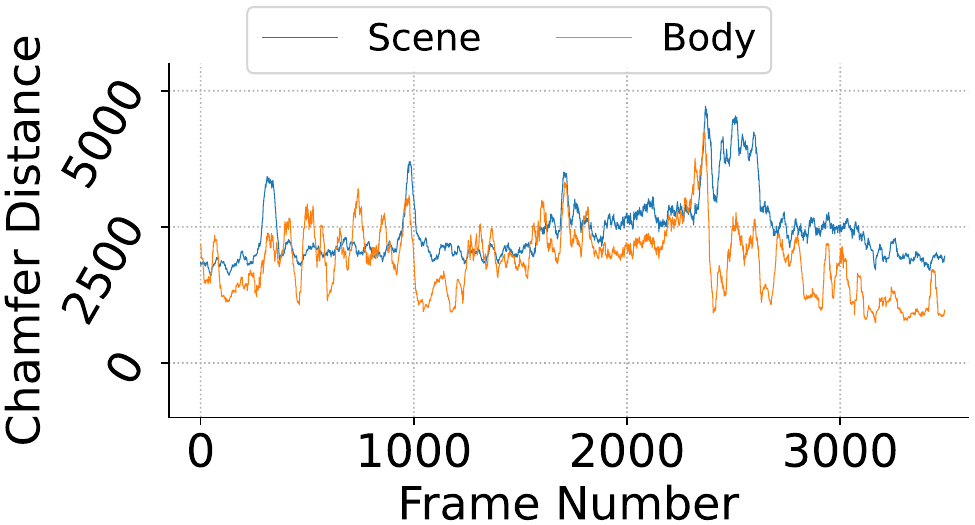}
        % \vspace{-0.5cm}
        \caption{Differences between consecutive 3520 frames to frame $N-1$ in a scene}
        \label{fig:measure-scene-change}
    \end{minipage}
     \hfill
    \begin{minipage}[t]{0.33\textwidth}
        \centering
        \includegraphics[width=0.9\columnwidth]{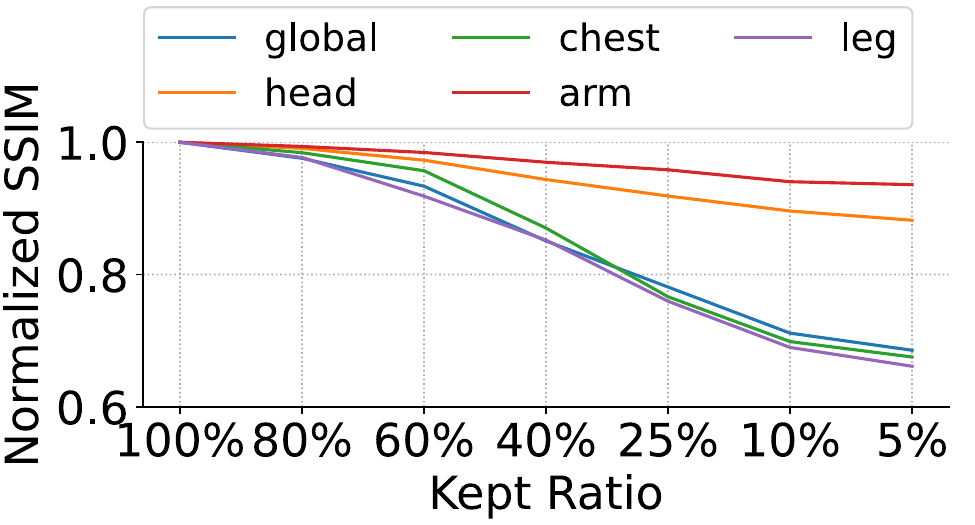}
        % \vspace{-0.5cm}
        \caption{Avg. Multi-view normalized SSIM of 4 body parts at increasing downsample rate}
        \label{fig:measure-visual-quality}
    \end{minipage}
\end{figure*}
%1.5-pages
\subsection{Measurement and Opportunity}

%Measurement要不要换成observation

\noindent\textbf{Dynamic and Static Volumetric Contents}
We first investigate the required bandwidth for live volumetric video streaming.
%测试传输全场景VV的bandwidth：
We measure the bandwidth consumption of transmitting full scene volumetric videos, using three pieces of volumetric video data of point cloud format selected from an open-source dataset FSVVD.
%测试每个part对bandwidth的contribution
%比较静态scene/object以及human body分别对bandwidth消耗了多少
As the orange bars show in Figure~\ref{fig: bandwidth_consumption}, we found that transmitting the full-scene volumetric video necessitates a bandwidth exceeding 10 Gbps. 
However, in real-world practice, the human body part in volumetric video contributes to the most visual changes, while the external scene remains static for most of the time. To quantify the bandwidth contribution of each part in the volumetric scene, we segment the volumetric scene into two parts: the dynamic human body and the static external scene, followed by transmitting them separately. The results are shown in blue and green bars, 
% \todo{explain octree-based method applied on the whole scene or static part only? 1) Use octree to determine the boundary of the static/dynamic part? 2) Use Octree to update the static part?}
which indicate: \textit{ The static external scene, on average, accounts for 92.8\% of the total bandwidth consumption, which significantly surpasses the bandwidth consumption of the dynamic human body at 7.2\%.} %From which we see a high potential in optimizing the bandwidth consumption by transmitting the VV part by part separately.

% \begin{figure}[t]
%     \centering
%     \includegraphics[width=0.9\columnwidth]{figures/measurements/measure-scene-change.png}
%     % \vspace{-0.5cm}
%     \caption{CD loss of the static scene on 4 video pieces \todo{Data.}}
%     \label{fig:measure-scene-change}
%     % \vspace{-0.5cm}
% \end{figure}

% \begin{figure}[t]
%     \centering
%     \includesvg[width=0.9\columnwidth]{figures/measurements/measure-visual-quality.svg}
%     % \vspace{-0.5cm}
%     \caption{Avg. SSIM of 5 body parts at increasing downsample rate.  \todo{Data incorrect.}}
%     \label{fig:measure-visual-quality}
%     % \vspace{-0.5cm}
% \end{figure}

%measurement:论证静态场景重复度高，以此来证明 hybrid scene 的必要性
%论证前后多个帧的重复度，把人拆分出来
\noindent\textbf{Scene Content Disparity}
According to the common sense that external scenes and objects in volumetric scenes are static most of the time, we analyze the disparity of the external contents to seek optimization opportunities. We exploit an Octree-based point cloud detection mechanism~\cite{Octree-based_detection} to compare the changes between adjacent volumetric video frames. In this measurement, we first convert the point cloud data into octrees, then traverse the octrees in a depth-first manner while comparing corresponding cells. Based on the comparison results, we quantify the difference using Chamfer distance~\cite{Chamfer_Distance_NIPS} and then obtain the results by aggregating the differences. We use the same three different scenes and show the results in Figure~\ref{fig:measure-scene-change}. We observed that: \textit{External scenes in volumetric videos are static in the playback process for most of the time, and changes that are visible to the naked eye are usually only located in specific regions that contribute small to the whole scene}.

%比较4个场景和1个完全静态的场景的 repeatability difference
%Coding to be updated

%(上面的内容可以画在一个图上, downsample to each metric, and compare the visual quality)

%测试decimation（downsample）之后图像质量的影响
%用不用解释一下我们为什么要这么做（测试decimation后的质量变化）？
\noindent\textbf{Visual Quality}
Along with bandwidth consumption measurement, we also assess the visual quality of volumetric content following PtCl decimation. In Figure~\ref{fig:measure-visual-quality}, we compare the visual quality of the same volumetric content at different point cloud density levels (PDL) to evaluate the impact of PDL decimation. We perform separate segmentation of the body parts and evaluate the quality of each part individually. Since there are no universally defined metrics for 3D visual quality assessment, we adopt SSIM, a commonly used metric in measuring 2D video quality. In the measurement process, we define a set of views and compare the SSIM of the 2D rendering of the 3D scene and the original 2D frame captured at the same predefined views. On average, we select 30 frames for each scene and compute the average SSIM for evaluation. The measurement encompasses the comparison of six downsampling rate levels with the original data, and we present the results using normalized SSIM. As shown in Figure~\ref{fig:measure-visual-quality}, we have the following observations: \textit{1) The visual quality has less than 11\% drop even after 50\% decimation in PDL. 2) Decimation degrades different quality to the different body parts (e.g., at 40\% downsample rate, head SSIM lost 3.1\% compared to 14.8\% for arm.)}

\begin{figure}[t]
    \centering
    \includegraphics[width=.8\columnwidth]{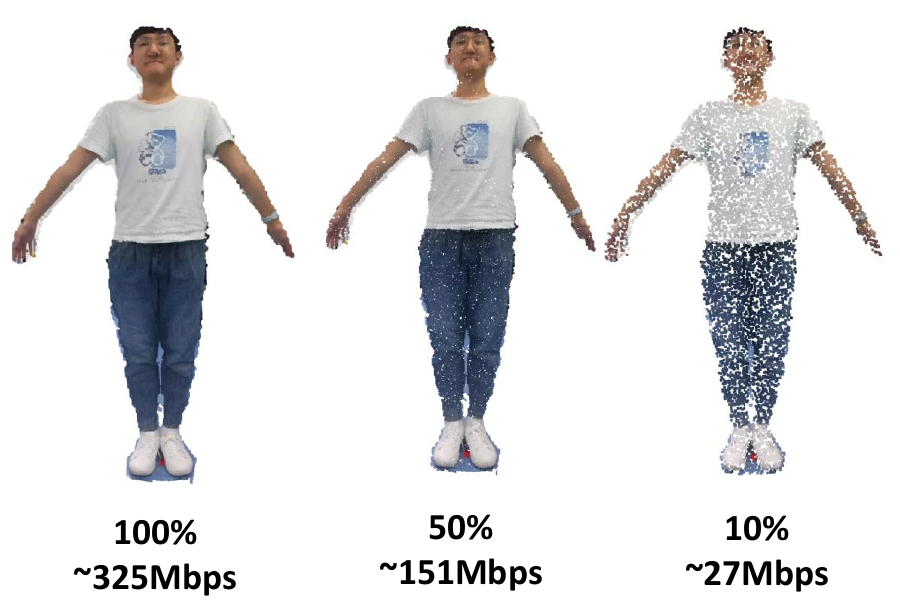}
    % \vspace{-0.5cm}
    \caption{Comparison of different PDL}
    \label{fig: different_PDL}
\end{figure}

\begin{figure}[t]
  \centering
  \begin{tabular}{ccc}
    \subfloat[Scene1]{\includegraphics[width=0.27\columnwidth,trim=30 0 30 0]{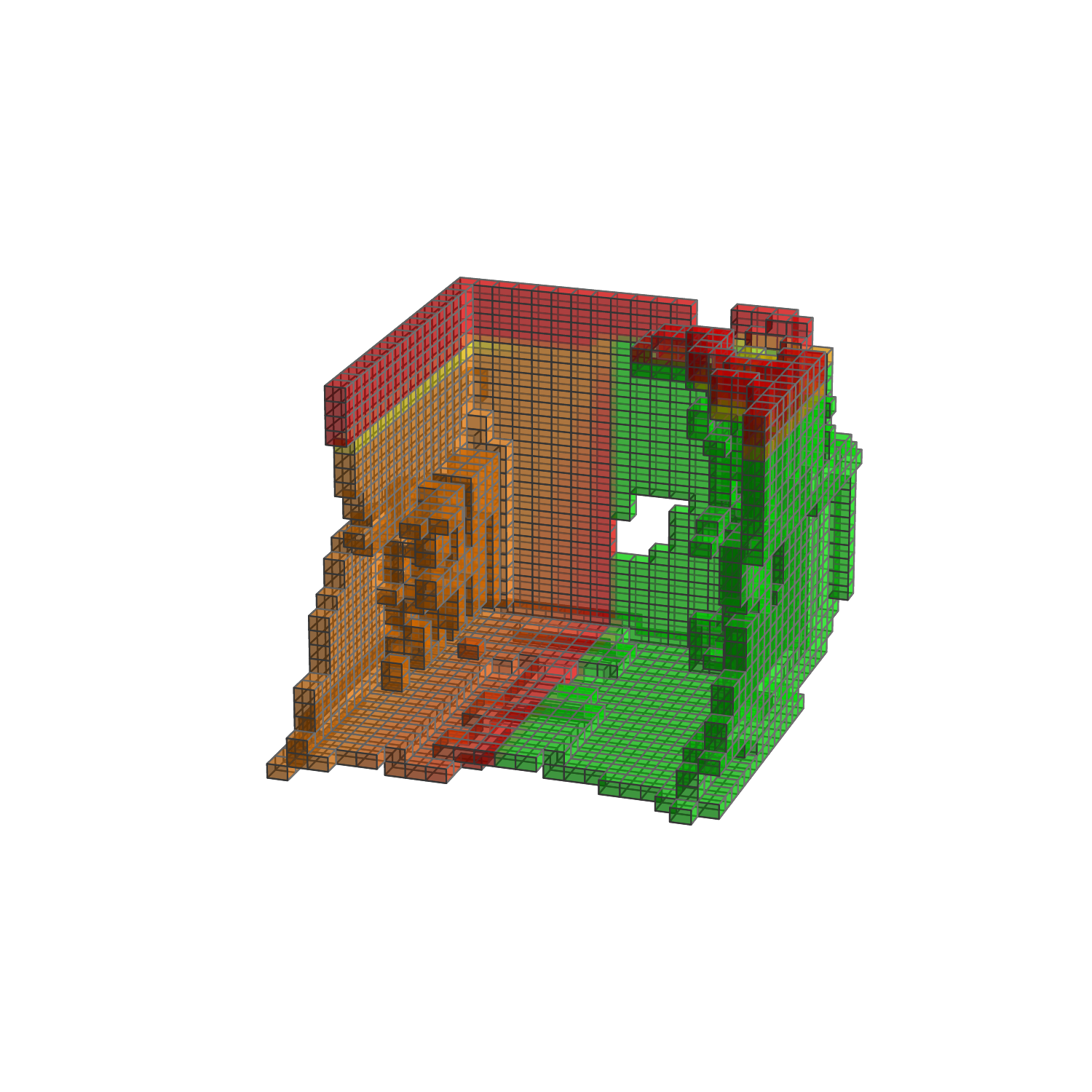}} &
    \subfloat[Scene2]{\includegraphics[width=0.27\columnwidth,trim=30 0 30 0]{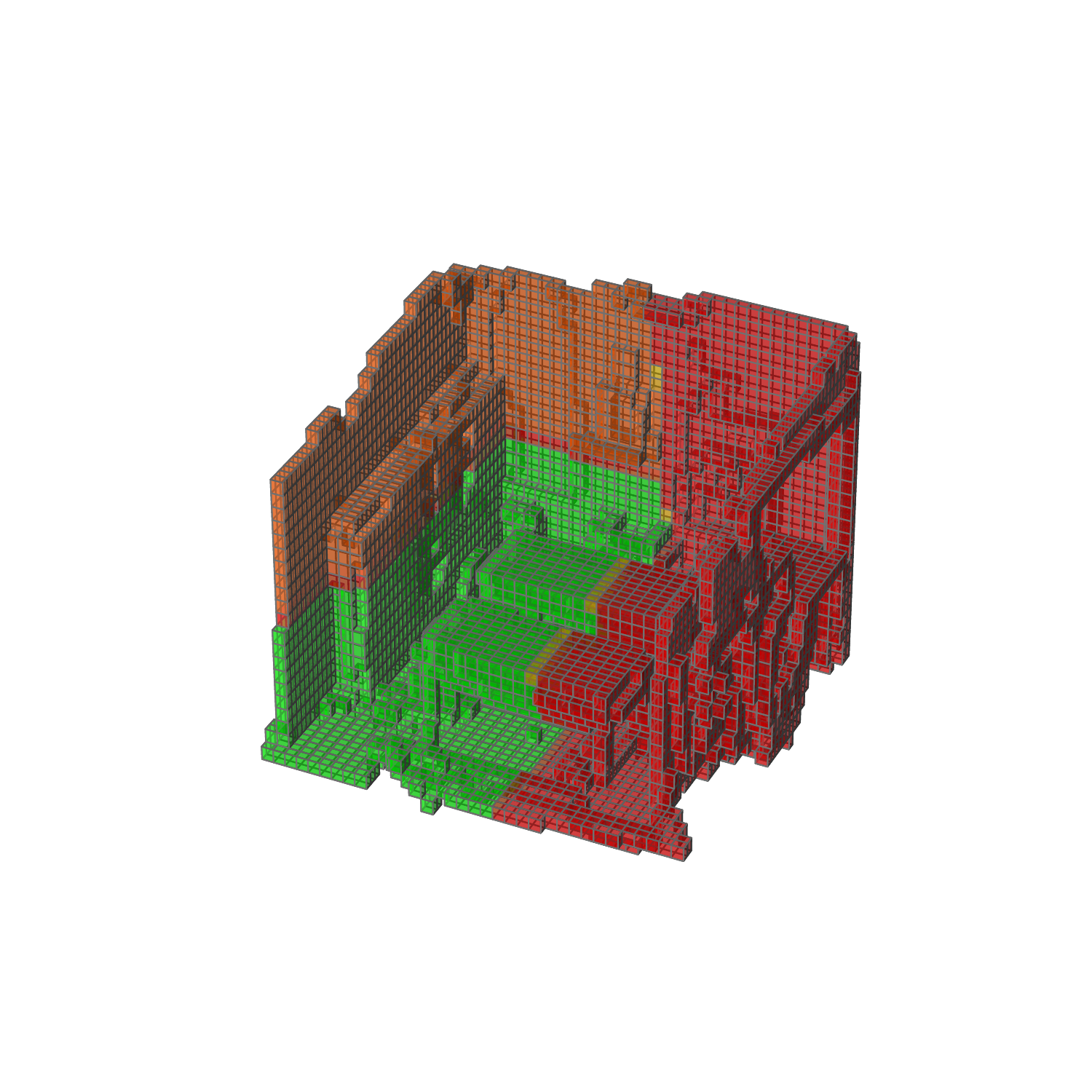}} &
    \subfloat[Scene3]{\includegraphics[width=0.27\columnwidth,trim=30 0 30 0]{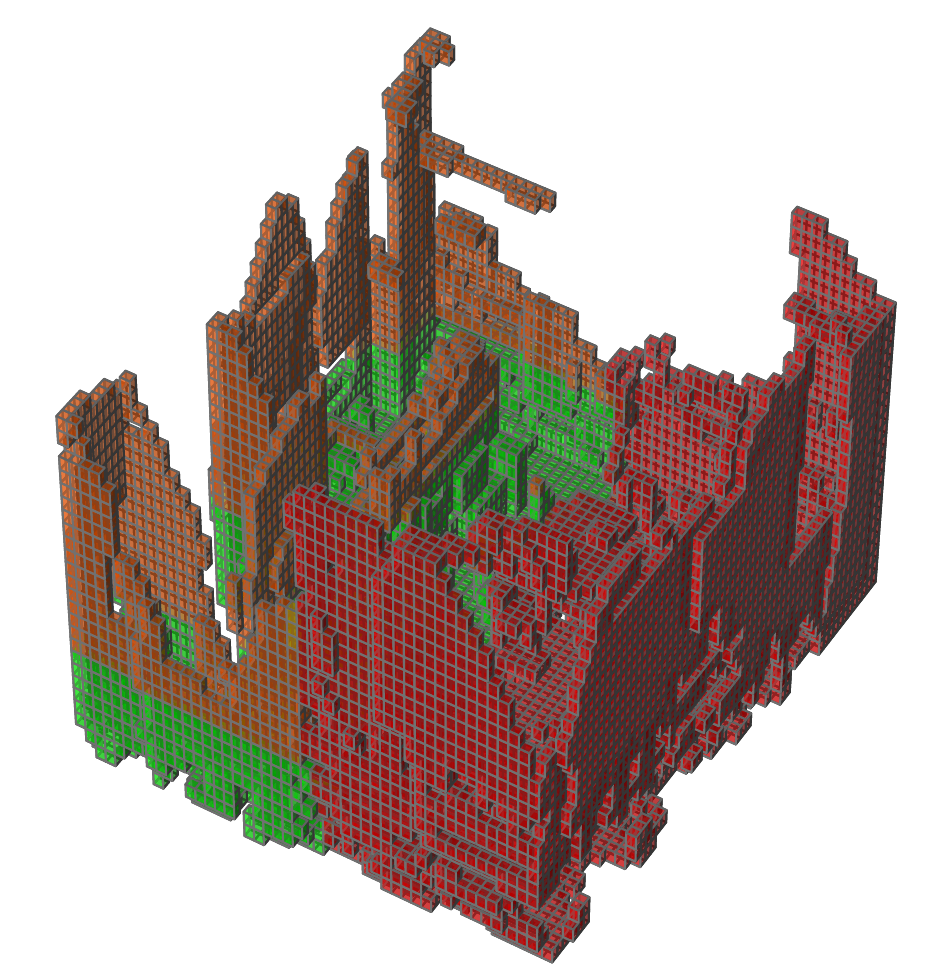}} 

  \end{tabular}
  \caption{Visualization of \textbf{Visual Saliency} in different scenes}
  \label{fig: Visual_Saliency_visualization}
\end{figure}

\noindent\textbf{Visual Saliency} Visual saliency is an important factor in the human visual system, that indicates the distribution of human attention in regions of interest (ROI) for a particular scene. Research on users' ROI for panoramic videos has been conducted in-depth and relevant results have been applied in practice~\cite{FoV_aware_360,Optimizing_360video_cellular}. But for volumetric video, related research is still in its infancy. We believe the users' viewing behavior can indicate the saliency of different regions in the volumetric video scene, providing insights for streaming optimization. 

For instance, we divide the whole volumetric scene into small cubes with side lengths of 15cm to perform the measurement and visualize the results with different colors according to the Visual Saliency. The results of 10 users' average visual saliency levels in three different scenes are given in Figure~\ref{fig: Visual_Saliency_visualization}, indicating: \textit{Users' Visual Saliency diverse from different volumetric video content; and the ROI are different in different scenes, with only a small portion of cubes exhibiting relatively high saliency levels.}

\begin{figure*}[t]
    \centering
    \includegraphics[width=\textwidth]{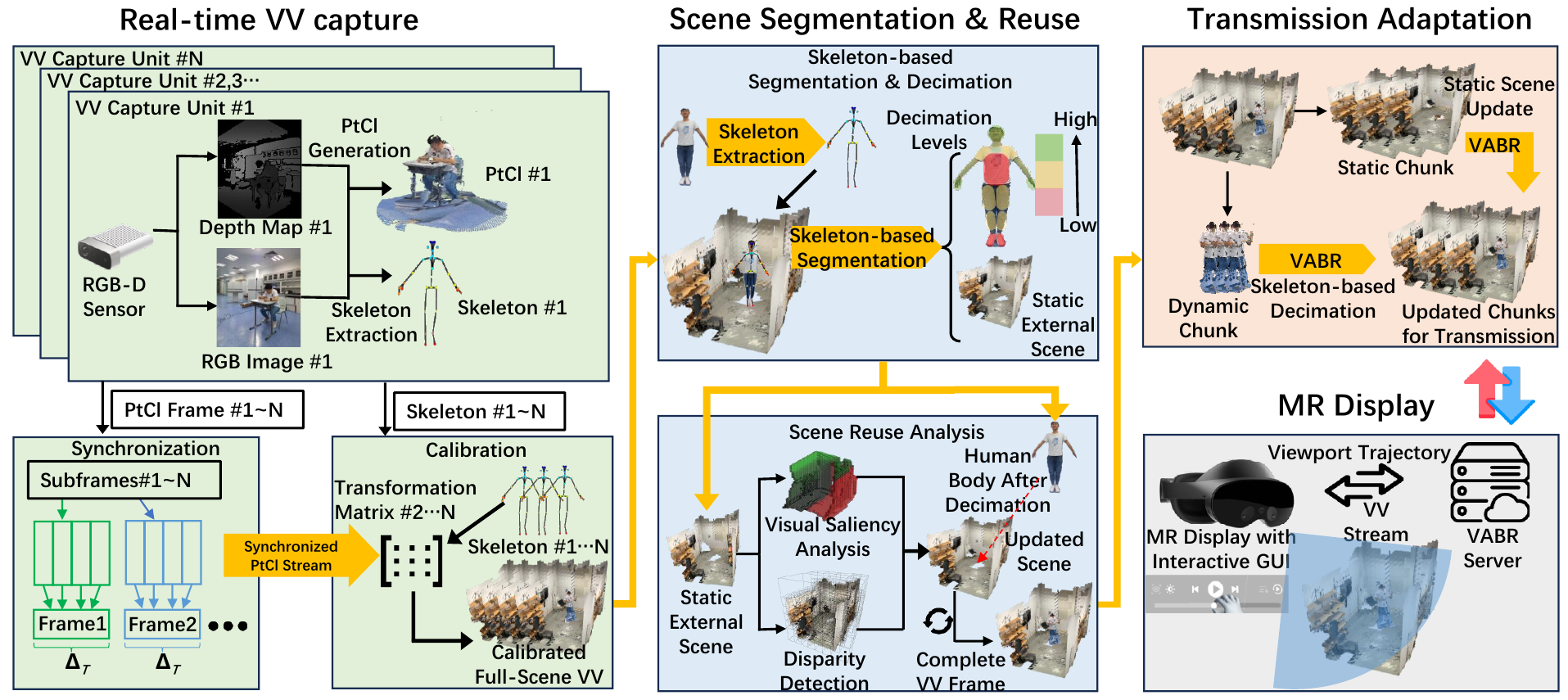}
    \caption{System Architecture of \textbf{LiveVV}}
    \label{fig: system_architecture}
\end{figure*}

\subsection{Motivation}
\label{sec: Motivation}
The above observations regarding bandwidth consumption, scene repetition, visual quality, and visual saliency provide valuable insights into our goal of achieving live volumetric video streaming:

\noindent$\bullet$\textit{Decimation to regions with low visual saliency offers a bandwidth-saving opportunity without introducing a noticeable visual quality drop.}

\noindent$\bullet$\textit{Leveraging already loaded static data with low disparity contributes to even greater bandwidth savings.}

\iffalse
%observations by measurement 1 & 2
\noindent$\bullet$\textit{Transmitting the entire volumetric content frame by frame directly would lead to a significant waste of bandwidth due to the limited disparity observed in static scenes, which contributes significantly to the overall data size.} 
%observations by measurement 3 & 4

\noindent$\bullet$\textit{Considering that certain regions within the volumetric content have a high tolerance for decreased point cloud density level (PDL) and many parts exhibit relatively low saliency due to user preferences, the visual quality impact of decimation on specific regions is acceptable. This allows for a more efficient allocation of bandwidth resources without compromising the overall user experience.}
\fi
These insights further motivate us to propose LiveVV to achieve live volumetric video streaming by optimizing the streaming process based on PtCl, i.e., instead of directly transmitting the entire volumetric scene repeatedly, we divide the volumetric scene into two categories: dynamic content with high disparity and static content with a low degree of disparity. In the streaming process, the dynamic content is continuously updated with careful decimation, while the static content is reused and updated according to need. We illustrate the design of the LiveVV in detail in the following Section~\ref{sec: system_design}, regarding the four modules: \textit{Real-time VV Generation}, \textit{Scene Segmentation \& Reuse}, \textit{Transmission Adaptation}, and \textit{MR Display}.

\section{System Design}
\label{sec: system_design}
%3-pages
The LiveVV is a holistic live volumetric video streaming system that encompasses the entire streaming pipeline. It seamlessly integrates real-time capture, scene segmentation \& reuse, transmission adaptation, and client display on MR headsets, achieving a frame rate of more than 24 fps with a latency of less than 350ms. The system is composed of four modules: Real-time VV Generation, Scene Segmentation \& Reuse, VV Transmission Adaptation, and MR Display.  We demonstrate the architecture of LiveVV in Figure~\ref{fig: system_architecture}. 

\subsection{Real-time VV Generation}
%介绍我们为啥要用人体追踪的calibration而不是meta stream里面的移动相机ORB calibration 方法

Previous studies have proposed several methods for capturing volumetric videos. Regrettably, these approaches have been unable to achieve real-time data output and have entailed intricate pre-capture configurations, thereby rendering them inapplicable in live-streaming scenarios. In light of these limitations, we hereby outline the challenges inherent in the development of a live volumetric video capture system as follows:

\noindent$\bullet$ The process of generating volumetric video entails complex operations, including the capture of RGB and depth frames, the mapping of these frames to create RGB-D frames and the subsequent generation of a colored point cloud stream. These steps impose significant computational demands and are highly sensitive to latency considerations, necessitating on-camera processing.

\noindent$\bullet$ Volumetric video capture units for live streaming scenarios require light and quick deployment due to the frequent change of capture region and fast movements of objects, thus making previous capture setups not applicable.

To fill this gap, we design a real-time volumetric video generation system based on a distributed architecture. By distributing the computing load to edge processing devices, and exploiting a body-track-based calibration method. This system is capable of capturing and delivering a full-scene volumetric video stream in point cloud (PtCl) at a high frame rate of up to 24 frames per second (fps). What sets the system apart is its ability to be deployed instantly without requiring any prior setups. The architecture of the capture system is illustrated in Figure~\ref{fig: system_architecture}. 

% \todo{Is the decision of using which method important? Should we categorize some methods and describe how we decide which to use?}

\subsubsection{Point Cloud Generation}
% \todo{Describe the process of constructing Point Cloud from an RGB-D frame with formulas}
% We reconstruct a point cloud for each frame from the captured RGB-D frame. \todo{Paraphrase. Copied from MetaStream} Given the configuration of cameras, we obtain the horizontal FoV ($h_{fov}$), the vertical FoV ($v_{fov}$), and the width ($d_{width}$) and height ($d_{height}$) of the original depth frame. Given a point ($x,y$) on the depth frame with depth value $z$, we calculate its point cloud coordinate ($x_{pc}, y_{pc}, z_{pc}$) values as: 

% \begin{equation}
% x_{pc} = dx \cdot \tan(\frac{h_{fov}}{2}) \cdot z \\
% y_{pc} = dy \cdot \tan(\frac{v_{fov}}{2}) \cdot z \\
% z_{pc} = z \\
% \end{equation}

% where $dx = 2 \cdot (x-\frac{d_{width}}{2})$

%\todo{describe the process of getting RGBD image from camera, and change the formula to more mathematical expression. }
For point cloud generation, the camera initially captures both RGB and depth images. We determine subpixel coordinates in the color image for each depth pixel, corresponding to the camera's depth mode. By employing bilinear interpolation, we extract precise color values, resulting in the generation of an RGBD image for subsequent point cloud generation.

Given the intrinsic and extrinsic parameters, we embark upon the computation of essential camera attributes, including the derivation of the camera pose, focal length, and principal point. Subsequently, leveraging the dimensions of the depth component within the RGBD image, from its height and width, we establish the appropriate sizing for both the points and their associated colors within the resultant point cloud. The spatial coordinates of the points within the point cloud are then determined as the following expressions:

\begin{equation}
z = D \cdot S \
x = \frac{(j - p_x) \cdot z}{f_x} \
y = \frac{(i - p_y) \cdot z}{f_y}
\end{equation}

where $D$ is the depth scale,
$S$ is the depth value at the pixel,
$j$ denotes the pixel's horizontal position,
$i$ signifies the pixel's vertical position,
$p_x$ is the principal point in the x-direction,
$p_y$ is the principal point in the y-direction,
$f_x$ is the focal length in the x-direction,
$f_y$ is the focal length in the y-direction.
Finally, the points are imbued with color information, thereby concluding the process of point cloud generation.

\subsubsection{Skeleton Extraction}
Motivated by prior research on body tracking~\cite{Easy_to_calibrate}, we develop a practical approach that utilizes skeleton joints as calibration markers. Our method employs a depth-based body skeleton extraction technique which involves the extraction of a skeletal representation from each frame. The representation consists of 32 distinct joints, each accompanied by its respective confidence level assessment. Following the judicious elimination of joints characterized by low confidence levels through a denoising process, we proceed to establish transformation matrices that facilitate the seamless integration of disparate coordinate systems. This is achieved by projecting sub-skeletons onto a master skeleton, thereby rendering the fusion of point clouds captured by multiple cameras a practicable endeavor within the context of the volumetric video generation system.

\subsubsection{Temporal Synchronization}
Temporal synchronization is a crucial process in constructing full-scene volumetric video frames by combining colored point clouds generated from capture units in different angles. Existing works primarily use hardware-based setups to trigger the depth cameras to take the shot simultaneously, which requires cable connections between devices~\cite{CWIPC_SXR,Holoportation}. As a result, such setups severely limit the mobility of the capture devices. 

In our design, we use a descriptor to encapsulate each volumetric video frame, enabling temporal synchronization. Before the data is sent out from each capture unit, each volumetric video frame is attached with a unique descriptor that contains the temporal information of this frame, indicating the precise time the frame is captured. Once the transmission server receives a frame, it places the frame into the same group for further processing, utilizing a pre-defined time slot $\Delta_T$. The number of frames in each time slot should correspond to the number of capture units, and the result frame rate is $\frac{1}{\Delta_T}$. In cases where frames are lost within a time slot, the previous frame is reused to prevent gaps in the scene for the current frame. In such a way, we get a synchronized complete frame sequence from multiple devices.

\subsubsection{Dynamic Camera Calibration}
The next step is to calibrate the frames of different angles. The result of the camera calibration is typically represented by a 4x4 transformation matrix, that describes the transformation that maps the generated point cloud data from the calibrated (`sub') camera to the reference (`master') camera. It encompasses both the intrinsic and extrinsic parameters obtained during calibration and enables the alignment of the point cloud data between different devices within the camera array. However, existing methods rely on additional settings, significantly restricting the mobility and practicality of the entire streaming system. Most existing volumetric capture methods require complicated and heavy deployment prior to capture, for example, CWIPC~\cite{CWIPC_SXR} requires placing a reference target to complete the calibration prior to capture, while FSVVD exploited a physical-structure-based calibration method that involves four boxes of specific size. Additionally, the conventional ORB calibration, exemplified by~\cite{ORB}, typically employs a chessboard pattern as the fiducial marker, which proves insufficient for accommodating scenarios that go beyond the spatial boundaries outlined by the marker. As a result, when merging frames captured by different cameras in the array, calibration-related inaccuracies inevitably arise, leading to suboptimal output. 

In order to achieve an anytime and anywhere deployment, we propose a dynamic camera calibration method that allows the capture units to be deployed at any time without the need for upfront preparation. Instead of extracting and mapping the features of reference substance prior to capture, we exploit the body track data to achieve calibration. Using the skeleton data received from the capture units as a reference, we derive the transformation matrix that maps the skeleton joints array from each `sub' frame to the `master' frame. Since the skeleton joints share the same coordinate system with the generated point clouds, the transformation matrix can be exploited to calibrate the PtCl frames from each angle as well. For a capture system with \textit{N} devices, the calibration involves \textit{N-1} matrix, that maps \textit{N-1} `sub' frames to the `master' frame.

By relying on the body track data and the derived transformation matrix, our calibration method enables seamless integration of the capture units without the need for traditional reference markers or pre-capture feature extraction. This approach enhances the versatility and ease of deployment for the streaming system, allowing it to be set up quickly and efficiently in various scenarios. Furthermore, to deal with the unpredictable movements of the capture units, the update frequency of transformation matrixes can be adjusted according to needs.

\subsection{Scene Segmentation and Reuse}

To prepare the captured volumetric video for subsequent transmission, we perform pre-processing on the captured data for optimization. According to the observations from Section~\ref{sec: Motivation}, we believe that applying partial decimation to the human body and updating corresponding external scenes instead of transmission frame by frame in live volumetric video streaming scenarios can significantly reduce bandwidth consumption while preserving the user's visual quality. To achieve that, we design a volumetric scene segmentation \& reuse mechanism that modifies the volumetric video frame prior to transmission. The component of the scene segmentation \& reuse mechanism is demonstrated in Figure~\ref{fig: system_architecture}. It consists of two parts: skeleton-based segmentation \& decimation and scene reuse analysis. 
%介绍content fusion都干了什么：
%The content fusion server first synchronizes the frames from each capture unit by combining them into the same time slot according to the attached time stamp of each frame, then 

\iffalse
\begin{figure}[t]
    \centering
    \includegraphics[width=\columnwidth]{figures/Content_Fusion_Server/skeleton_filter.pdf}
    % \vspace{-0.5cm}
    \caption{Skeleton-based Segmentation \& Decimation}
    \label{fig: skeleton_filter}
    %\vspace{-0.5cm}
\end{figure}
\fi

\iffalse
The conventional employment of ORB calibration engenders restrictive implications for VV content creation, as it confines volumetric capture to spatially delimited regions, primarily contingent upon the placement of fiducial markers, e.g., a chessboard pattern. This paradigmatic approach inherently imposes constraints, as it fails to address scenarios beyond the marker-defined confines. Consequently, when frames acquired from distinct cameras within the array are amalgamated, calibration-induced inaccuracies invariably manifest, leading to suboptimal output. Embracing body-tracking-based calibration offers a transformative solution, liberating users from marker-defined constraints and enriching their immersive interactions with the VV streaming system. 
\fi

%先说人（dynamic）部分的传输再说static的传输
\subsubsection{Skeleton-based Segmentation}

%Segment the point cloud of the human body based on the captured skeleton data.
Inspired by the observations that \textit{the human body part only contributes 7.2\% to the total bandwidth consumption during the streaming process}, we first propose a skeleton-based filter to segment the dynamic human body from the static external scene. By exploiting the results of body tracking, we design a skeleton-based filter that extracts the human body PtCl. During segmentation, only the points that fall into the filtering region are treated as human body PtCl and preserved. In addition, the filter performs segmentation by 15 cylindrical filters whose axes are defined by adjacent joints in the detected skeleton, that output the PtCl of each part of the body. After segmentation, we can label the result point cloud of each part, e.g. the PtCl of 'Head' or 'Right lower arm', as depicted in Figure~\ref{fig: system_architecture}.

\begin{figure}[t]
    \centering
    \includegraphics[width=\columnwidth]{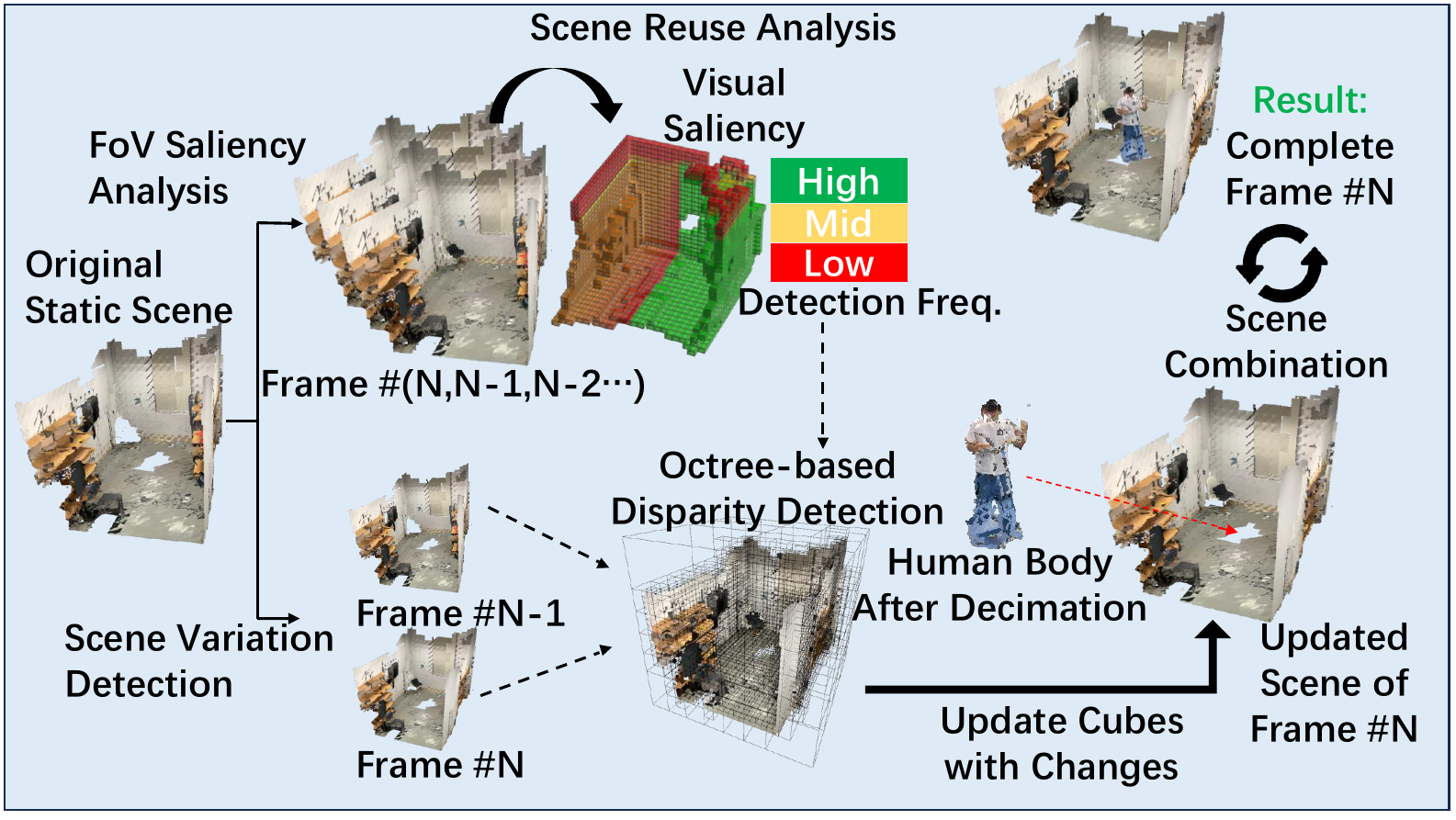}
    % \vspace{-0.5cm}
    \caption{Workflow of Scene Reuse Analysis}
    \label{fig: Hybrid_Scene}
    % \vspace{-0.5cm}
\end{figure}

\subsubsection{Scene Reuse Analysis}
\label{sec: Hybrid_Scene_Analysis}
%Octree-based point cloud detection. 
LiveVV manipulates the points of external objects individually and combines them with the extracted human body data to compose a complete volumetric scene. Such processing contributes much to bandwidth savings as the measurement results in Section~\ref{sec: measurement} indicate high repeatability of external objects in multiple different scenarios. To achieve this, we initially partition the external volumetric scene into smaller cubes. Subsequently, we analyze the visual saliency of each cube to determine its update frequency. By comparing the disparity level of each cube with its corresponding cube from previous instances, we selectively update the external scene at the cube level. This granular approach ensures that only the necessary changes are transmitted, further enhancing bandwidth savings.

\noindent\textbf{Visual Saliency Analysis}
Visual saliency is an important factor in the human visual system, indicating the distribution of regions of interest (ROI) in a particular scene. In this work, we consider the visual saliency when determining scene updating frequencies. Based on the divided cubes, we measure the visual saliency of each region by a quantitative indicator to determine the users' level of attention to each region, by exploiting the calculation Formula~\ref{formula: FoV_saliency}. 

Unlike traditional 2D videos, the visual saliency of volumetric videos can not be represented by a specific region in the frame plane due to their 3D nature. We hereby define an indicator as `Visual Saliency', \textit{$S_V$}, taking the following factors into consideration: PtCl density level (PDL), visual distance, and appearance frequency in viewing frustum (a truncation with two parallel planes of the pyramid of vision)~\cite{viewing_frustum} into consideration. The formulation of \textit{$S_V$} is given as: 

\begin{equation}
    \textbf{\textit{$S_V$}}=\frac{\rho_c*\textit{$f_g$}}{\textit{$D_c$}}
    \label{formula: FoV_saliency}
\end{equation}
%$$\textbf{\textit{$F_a$}}=\rho_c*\textit{$f_g$}$$
\begin{equation}
    \textit{$f_g$}=\frac{\sum_{i=1}^N\textit{$N_g($i$)$}}{\textit{$N_{sample}$}}
\end{equation}
%$$\textit{$f_g$}=\frac{\sum_{i=1}^N\textit{$N_g($i$)$}}{\textit{$\#Samples$}}$$
where \textbf{\textit{$\rho_c$}} is the point cloud density of the cube, \textbf{\textit{$f_g$}} is the frequency of each cube falling into the viewing frustum, \textbf{\textit{$D_c$}} is the distance between the headset and the cube center, \textit{$N_g($i$)$} represents the count of the current cube, if the cube falls within the viewing frustum in sample $k$, then $N_g(k)$ is set to 1, otherwise, it is set to 0, and \textbf{\textit{$N_{sample}$}} is the total number of headset tracking samples. 

To accurately track the users' movement and viewport trajectory, we employ data from the headset's built-in accelerometer and eye tracker. By aligning the coordinate system and analyzing this data, we gain insights into the users' current and historical movement patterns and their corresponding viewport trajectories within the volumetric scene. Typically, the rotational angles from the headset are represented using Euler angels, for the convenience of viewport calculation, we transform the angel into radians, where $\alpha$, $\beta$, and $\gamma$ stand for yaw, pitch, and roll movement respectively. We use a matrix \textit{R} to represent the rotation matrices of yaw, pitch, and roll. 
\begin{equation}\label{eq:rotation_matrix}
R = 
\begin{bmatrix}
    1 & 0 & 0 \\ 
    0 & \cos\gamma & -\sin\gamma \\ 
    0 & \sin\gamma & \cos\gamma
\end{bmatrix}
\begin{bmatrix}
     \cos\beta & 0 & \sin\beta \\ 
     0 & 1 & 0 \\ 
     -\sin\beta & 0 & \cos\beta 
\end{bmatrix}
\begin{bmatrix}
    \cos\alpha & \sin\alpha & 0 \\ 
    \sin\alpha & \cos\alpha & 0 \\ 
    0 & 0 & 1
\end{bmatrix}
\end{equation}

We can calculate the viewport vector based on the above matrix, then we form a direction vector formed by the coordinates of the headset and the center of each cube. By comparing the angle between the direction vector and the viewport vector, we can know whether the cube falls into the user's viewing frustum.
Normally, the naked eye can see only a very limited range of objects~\cite{eye_viewing_angle_1, eye_viewing_angle_2, eye_viewing_angle_3} within about 30°, thus we set the threshold angle as 30°. %具体介绍计算的规则
Cubes with the center falling into the viewing frustum are allocated with higher visual saliency.

We iterate all of the cubes in one frame to get a global saliency score, since the movement tracking frequency is larger than the live streaming frame rate, we combine the multiple FoV data together for an average saliency calculation. Based on the saliency analysis results, the scene disparity detection frequencies are mapped to three levels: low, mid, and high, as shown in Figure~\ref{fig: Hybrid_Scene}.

\noindent\textbf{Scene Disparity Detection}
%Detect the change in external objects using octree detection and update the changed objects.
We utilize an Octree-based detection mechanism to detect disparity in the extracted external scene. %具体怎么detect
Initially, we partition the volumetric scene into uniform cubes with adjustable side lengths for different detection granularity based on prevailing network conditions. Since the volumetric frames are aligned to a common coordinate system, each cube is assigned a unique coordinate to enable straightforward comparison with cubes at corresponding positions in adjacent frames.
%detection如何实现：
As the cubes in adjacent frames have the same size and similar densities, we exploit Chamfer Distance~\cite{Chamfer_Distance_NIPS} as the comparison metric.
To enhance the efficiency of calculation, we first build Octrees for the point clouds in each cube, the octree subdivides the volumetric contents into octants, such a method provides a hierarchical spatial data structure that can accelerate nearest-neighbor searches and reduce the computational complexity when comparing point cloud sets. Then we traverse the Octrees to perform a nearest neighbor search for each point, starting from the root octant and recursively traversing the Octree by examining child octants, using Euclidean distance to measure the spatial dissimilarity between the points. The calculation of Chamfer Distance ($C_d$) is as follows: 
\begin{equation}
    \text{\textit{$C_d$}}(P, Q) = \frac{1}{|P|} \sum_{p \in P} \min_{q \in Q} |p - q|_2 + \\\frac{1}{|Q|} \sum{q \in Q} \min_{p \in P} |q - p|_2
\end{equation}
where \textit{P} represents the first point cloud, consisting of $|P|$ points, \textit{Q} represents the second point cloud, consisting of $|Q|$ points, and $|\cdot|_2$ denotes the Euclidean distance between two points, which is computed as the square root of the sum of squared differences in each dimension.

\begin{figure}[t]
    \centering
    \includegraphics[width=\columnwidth]{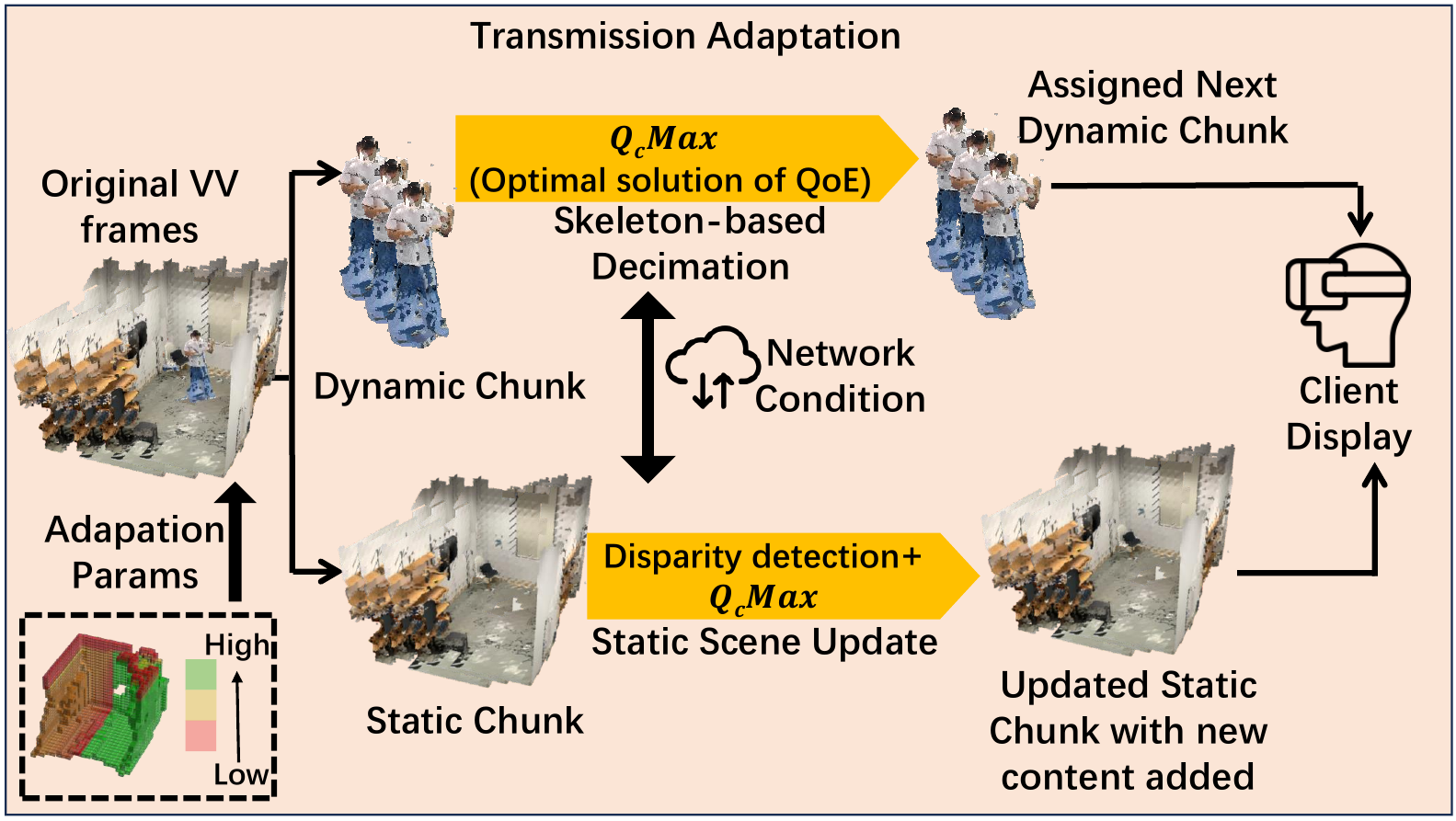}
    % \vspace{-0.5cm}
    \caption{Workflow of Transmission Adaptation}
    \label{fig: Transmission_adaptation}
    % \vspace{-0.5cm}
\end{figure}

\subsection{Transmission Adaptation}

%add one figure: comparison of VV and 2D video before and after down-sampling 75%

To meet the high bandwidth requirements of live volumetric video streaming and ensure a consistent quality of experience (QoE) for users, LiveVV incorporates a bandwidth adaptation mechanism, as illustrated in Figure~\ref{fig: Transmission_adaptation}. Unlike traditional 2D video streaming optimization, simply switching the resolution according to the bandwidth changes will have a huge impact on the visual quality. As illustrated in Figure~\ref{fig:measure-visual-quality}, there is a significant drop in visual quality after data decimation in volumetric data. To overcome such an issue, we propose the volumetric video adaptive bitrate streaming algorithm (VABR) to dynamically adjust the streaming parameters based on the available network bandwidth. VABR exploits a similar scheme as DASH (Dynamic Adaptive Streaming over HTTP). During transmission, the volumetric video stream is packaged into chunks, and the streaming parameters are adaptively adjusted to optimize the streaming quality according to available bandwidth.

% To enable real-time transmission, we define a binary file structure: .PCV (Point Cloud Video), which compacts multiple volumetric video frames into a single chunk. Such file structure deals with the constantly changing frame data size of volumetric videos, as opposed to the fixed pixel count in traditional videos. The structure of .PCV is shown in Figure \ref{}, the frame rate and frame count are first defined in the header for each chunk, allowing for efficient indexing and data access.
%介绍一下PCV的格式

\subsubsection{QoE for Volumetric Video}
The ultimate goal of VABR is to get the best QoE under fluctuating network conditions to provide a better experience. We define a few QoE metrics for volumetric video and combine them in one model to enable an adjustment to different preferences. We enumerate the following key metrics of volumetric video QoE:

\noindent$\bullet$ \textit{Average VV Quality:} The average visual quality in one volumetric video chunk, that can be measured by analyzing the 2D representations from users' viewpoint:
$\frac{1}{K}  {\textstyle \sum_{k=1}^{K}} q(R_k)$
, where $R_k$ is the bitrate of down-loading chunk \textit{k}

\noindent$\bullet$ \textit{Average Quality Variation:} The variations in QoE among different chunks:
$\frac{1}{K-1}  {\textstyle \sum_{k=1}^{K-1}} |q(R_{k+1})-q(R_k)|$

\noindent$\bullet$ \textit{Startup Delay:} $T_s$

We define the QoE of chunk 1 through \textit{K} by a weighted sum of the above four metrics due to users' different preferences:
%%single line version:
%\begin{equation}  
%   \textit{QoE}_{1}^{K}={\sum_{k=1}^{K}}q(R_k)-\lambda\sum_{k=1}^{K-1}|q(R_{k+1}-q(R_k)| 
%    -\mu \sum_{k=1}^{K}(\frac{d_k(R_k)}{C-K}-B_k)_+-\mu_sT_s
%\end{equation}

\begin{equation}
\textit{QoE}_{1}^{K} = \sum_{k=1}^{K} q(R_k) - \lambda \sum_{k=1}^{K-1} |q(R_{k+1}) - q(R_k)| - \mu_s T_s
\label{QoE_model}
\end{equation}

where $\lambda$ and $\mu_s$ are no-negative weighting parameters, corresponding to \textit{Average Quality Variation} and \textit{Startup Delay}. For example, a small $\lambda$ means the user is not sensitive to the variation of the visual quality, while a larger $\lambda$ means the user focuses more on the smoothness of the playback. While larger $\mu_s$ indicates the user prefers a lower startup delay.

\subsubsection{Design of VABR}
% \todo{With volumetric video chunk saved in .PCV. We can exploit the currently available rate adaptation algorithm for the PCV format without making many changes.} 
VABR stems from an existing optimization framework to achieve volumetric video bandwidth adaptation~\cite{DASH_SIGCOMM}. In practice, it models one volumetric video stream chunk into two distinct chunks, a dynamic chunk, and a static chunk. 

%\noindent\textbf{Layered Representation of VV}
In a traditional DASH scheme, a chunk comprises multiple independently encoded versions of the video, each catering to different quality levels. But for volumetric video, instead of directly performing decimation to the PtCl stream and encoding them into multiple independent chunks, VABR exploits the unique nature of PtCl by incrementally upgrading the quality of the layers. Since volumetric video frames in PtCl format are represented by hundreds of thousands of colored points that are unordered, we can adjust the quality by exploiting a layered representation of volumetric video:

%\noindent$\bullet$ For the dynamic human body, only the most important parts (the head part) are preserved with full-scale PDL in the lowest layer, while other parts only contain the lowest PDL that can represent the outline of the body. For higher layers, each part is enhanced by increasing the PDL at different portions. The highest layer r contains all of the points for the whole body.

\noindent$\bullet$In the case of \textbf{dynamic human body} representation, VABR employs a selective approach to preserve the most important parts, with the head being a prime example. These vital parts are retained with full-scale Point Cloud Density Level (PDL) in the lowest layer. On the other hand, the remaining body parts are represented with a relatively low PDL that can effectively outline the body structure. As we ascend to higher layers, each individual body part undergoes enhancement by increasing the PDL at specific regions. This increment in PDL allows for a more detailed and refined representation of each part. Finally, the highest layer encompasses all the points necessary to reconstruct the complete body, ensuring a comprehensive representation.

%\noindent$\bullet$ For external static scenes that are divided into cubes of uniform size, only the lowest layer contains the points to construct the scene, while higher layers will add more detail to each region of the scene. For each layer, the portion of each cube is determined by the history visual saliency score, regions that never appear in users' viewing frustum will not be updated instantly.

\noindent$\bullet$In the case of \textbf{external static scene} divided into uniform-sized cubes, VABR adopts a layer-based approach. In the lowest layer, only the points necessary to construct the scene are included. As we move to higher layers, additional details are progressively added to each region of the scene. The allocation of points within each layer's cubes is determined based on the historical visual saliency score. This ensures that regions of the scene that have greater visual significance or attract more user attention are prioritized for higher detail representation. Conversely, regions that consistently fall outside of users' viewing frustum and have minimal visual relevance may not be instantly updated.

Given that users' visual saliency is primarily focused on dynamic content, our approach prioritizes the transmission of dynamic chunks~\cite{hu2023understanding}.  During the transmission process, VABR periodically examines a finite window of the next N chunks for dynamic chunks and searches for the most suitable assignment that maximizes a utility function derived from the QoE model (Equation~\ref{QoE_model}). Considering the optimization for volumetric videos with hybrid scene composition, instead of fetching an independent chunk with a matching quality level for the next playback period, VABR upgrades each whole frame of dynamic chunks by incorporating the changes in the subsequent chunks. For example, for an already-fetched chunk with quality level $Q_c$, VABR will set the search space from $Q_c$ to max quality level $Q_cMAX$, if the result is a higher level than $Q_c$, then additional layers will be fetched to enhance the current chunk.  

On the other hand, for static chunks, the disparity of cubes is detected at a different frequency according to the saliency score as discussed in Section~\ref{sec: Hybrid_Scene_Analysis}. Only cubes with changes are updated by incorporating added data from the next chunk. Similar to the upgrading scheme for dynamic chunks, for static chunks, a chunk with a higher quality level will be fetched to enhance the current chunk under better network conditions to improve the quality.

\subsubsection{Skeleton-based Decimation}
To optimize bandwidth consumption during the transmission of dynamic human body data, we utilize a technique called partial point cloud decimation for each individual body part. The fundamental idea is to selectively apply decimation to the point cloud of various human body parts, tailoring the level of reduction to different PDLs.

In order to determine the appropriate downsampling proportion of each body part, we make the following measurement: Based on the segmentation result of the skeleton-based filter, we extract the point cloud data of each part individually. We then perform downsampling to each part at specific rates and then compare the impact on visual quality by analyzing the SSIM of the 2D rendering of the 3D volumetric parts, using the same method as described in Section~\ref{sec: measurement}. Based on the test results, we allocate different downsampling proportions to each body part, as depicted in Figure~\ref{fig: system_architecture}. The downsampling process is executed prior to the transmission, with real-time body tracking, we can perform the partial decimation for each frame of the human body in real-time. The bitrate of transmitting the dynamic chunk \textit{k} can be formulated as:
\begin{equation}
    R_d(k)= {\textstyle \sum_{f=1}^{N_f}} {\textstyle \sum_{i=1}^{15}} P_b(i)
\end{equation}
where $N_f$ is the frame number of the dynamic chunk \textit{k}, $R_d$ is the total bitrate of transmitting the current level of dynamic chunks, $P_b(i)$ is the point cloud density of body part \textit{i} after decimation in current network conditions.  
%, where $R_k$ is the optimal solution of Equation.~\ref{QoE_model}.
\subsubsection{Static Scene Update}
%combine scene variation and FoV saliency
We update the external static scene at the cube level, by integrating the results of visual saliency analysis and disparity detection. We first analyze the visual saliency to determine the detection frequency of each cube, as described in Section~\ref{sec: Hybrid_Scene_Analysis}. Subsequently, based on the visual saliency score, we establish three levels of disparity detection frequency: `High' (every frame per detection, `Mid' (5 frames per detection), and `Low' (10 frames per detection) as shown in Figure~\ref{fig: Hybrid_Scene}.

For each cube, the corresponding $C_d$ value is calculated along with its most recently updated cube at the allocated detection frequency. If a cube has no past cube then it is directly updated. On the other hand, if a previous existing cube is empty in the current frame, it will be replaced by an empty cube as well. The utilization of visual saliency ensures that objects within the user's FoV are promptly updated thus ensuring good visual quality. Conversely, objects outside the user's FoV will not be updated in each frame, thereby preventing unnecessary transmission bandwidth waste. Furthermore, in order to adapt the streaming process to the network fluctuations, we introduce variation updating quality, which also exploits VABR to determine the quality of chunks to be fetched. Since the dynamic chunks contribute more to the user's visual saliency, we prioritize the allocation of network bandwidth to dynamic chunks. Thus, the bitrate of updating static chunk \textit{k}, $R_s(k)$ can be formulated as follows:

\begin{equation}
    R_s(k) = R_k-R_d(k)
\end{equation}
\begin{equation}
  R_s(k)= {\textstyle \sum_{i=1}^{N_c}}\frac{C_d}{C_dmax} P_c(i)*F_d 
\end{equation}

\begin{equation}
F_d = \begin{cases}
  & 1\text{ if } S_v \ge  16.0\\
  & 0.2\text{ if } S_v \ge  9.0\\
  & 0.1\text{ if } S_v < 9.0
\end{cases}  
\end{equation}
where $P_c(i)$ is the point cloud density of cube \textit{i}, and we use a normalized chamfer distance to determine the PDL for updating, $N_c$ is the total number of cubes needing to be updated, $F_d$ is the disparity detection frequency, which is determined by the visual saliency score $S_v$.

Given the optimal solution of the QoE model, we can get the remaining available bandwidth $R_S$, for updating the static scene.
\subsection{MR Display}
To deliver a truly immersive experience to users, we develop a client-side Mixed Reality (MR) player. This player is specifically designed to provide seamless integration and interaction with volumetric video streams. It receives the video stream from the VABR server, which dynamically adapts the transmission based on network conditions. Additionally, the client-side MR player continuously tracks and reports the user's viewport trajectory. This information is invaluable for assisting visual saliency analysis on the VABR server, enabling optimized streaming and rendering of the most relevant content. The entire workflow is visually depicted in Figure~\ref{fig: system_architecture}, showcasing the interconnectedness of the components. To further amplify interactivity, we have integrated a virtual graphical user interface (GUI) directly into the MR player. This GUI grants users direct access to several additional functions, enhancing their ability to engage and interact within the immersive environment:

\noindent$\bullet$ A content management system that is capable of adding external objects in the volumetric scene.

\noindent$\bullet$ A `Passthrough Mode' that allows the user to switch from a fully immersive virtual scene to a mixed reality scene that includes both real and virtual scenes.

\noindent$\bullet$ An ambient light switch to adjust the illumination in the virtual scene.
\iffalse
\begin{itemize}
    \item A content management system that is capable of adding external objects in the volumetric scene.
    \item A `Passthrough Mode' that allows the user to switch from a fully immersive virtual scene to a mixed reality scene that includes both real and virtual scenes.
    \item A ambient light switch to adjust the illumination in the virtual scene.

\end{itemize}
\fi

\begin{figure}[t]
    \centering
    \includegraphics[width=\linewidth]{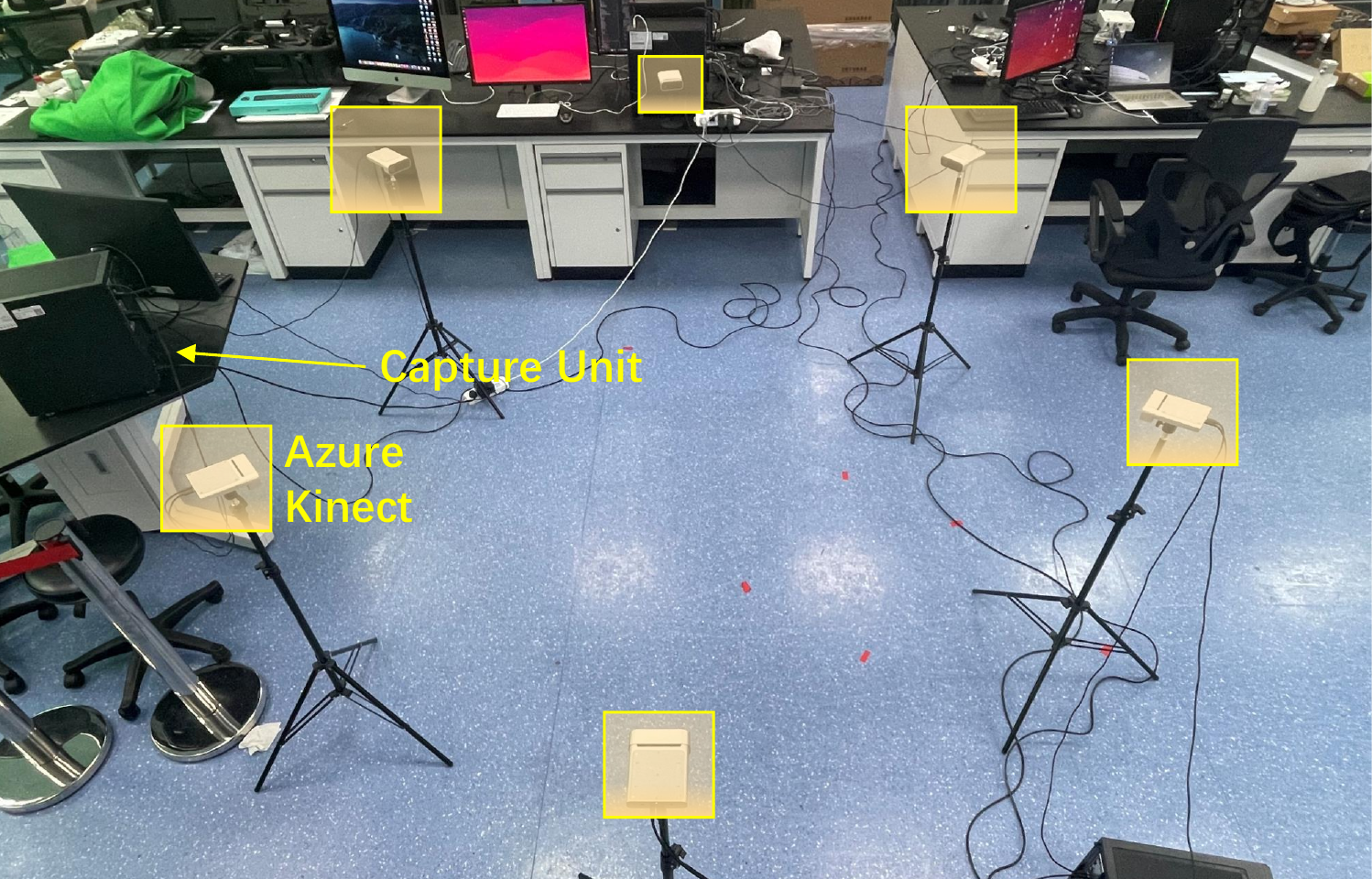}
    \caption{System Implementation of \textbf{LiveVV}}
    \label{fig: System_implementation}
\end{figure}

\section{System Implementation}
\label{sec: System_implementation}
%0.5-pages
\head{Hardware} We implement the system using commodity devices. The hardware setup includes three capture units, each equipped with an Intel i7-12700 CPU and attached with two Azure Kinect depth sensors. %Four content fusion servers each with two 10-core Intel Xeon Silver 4210 CPUs, 64GB RAM, and one of them is equipped with an NVIDIA RTX 3090 GPU. 
One VABR server with two 10-core Intel Xeon Silver 4210 CPUs, 64GB RAM, and an NVIDIA RTX 3090 GPU. They are interconnected with LAN. The MR client utilized a Meta Quest Pro headset. The implementation setups of the system are shown in Figure~\ref{fig: System_implementation}.

\head{Software} The software implementation of LiveVV is developed in Python and C++. In our experimental implementation, we employ six Azure Kinect cameras to construct an array for capturing full-scene volumetric video. We capture raw color images and depth maps from the camera at a rate of 30 FPS. The resolution is set to 1280x720 for color images and 320x288 for depth maps (utilizing the NFOV 2X2 binned mode of the depth sensor to achieve the highest depth resolution). We utilize Open3D~\cite{Open3DZhou2018} to combine the color and depth data into RGB-D images, which are then converted into colored point cloud streams. We implement the VABR server with transmission adaptation capabilities: it receives point cloud frames from the capture units, synchronizes the frames and merges them based on the skeleton-based calibration results. Then, the merged PtCl stream is segmented into two parts: static external scene and dynamic human body, which are further manipulated to separate chunks for adaptive transmission. The MR client is developed in Unity, allowing users to manipulate the volumetric video frame during live streaming.

\section{Evaluation}
In this section, we evaluate the performance of LiveVV in the following aspects: frame rate, visual quality, latency, and bandwidth consumption, including two tracks: the performance of each module and the overall performance of the whole system. %We set up the system in a room, live capture, and process for two minutes (3600 frames), with one person entering the camera circle, pushing a box or a chair across it, and leaving it several times. 
We implement the system using the setups as described in Section~\ref{sec: System_implementation} and present the result using FPS (Frames Per Second), SSIM (Structure Similarity Index Measure), end-to-end latency, and network bandwidth consumption as the measurement metrics.

\iffalse
\begin{itemize}
    \item \textbf{Practicality.} What is the bandwidth and computation power saving that each step of our system has contributed? % What is the bandwidth and computation usage of each proposed step for LiveVV to stream in \textit{real-time}? 
    \item \textbf{Visual Quality.} What is the visual impact of our skeleton-based dynamic-static content decimation mechanism?
\end{itemize}
\fi

% To evaluate the performance of LiveVV in practical application scenarios, we conduct several assessments in the aspects of: VV capture quality, bandwidth consumption, 

% \todo{Key message: (1) our system is practical: can do each of the proposed steps in real-time; and can transmit in reasonable bandwidth; (2) our visual quality is good: we quantitatively measure the results of the output stream.}

\subsection{Skeleton-based Segmentation and Decimation}

\begin{table*}[t]
    \centering
    \begin{minipage}{.45\textwidth}
        \centering
        \begin{tabular}{c|c|c|c|c|C{1.3cm}|C{1.4cm}}
            \toprule
             & \multicolumn{4}{c|}{Kept Ratio (\%)}& Bandwidth & Normalized\\ \cline{2-5}
             \#&  head&  chest&  arm&  leg& (Mbps) & SSIM\\ \midrule
             base &  100&  100&  100&  100& 325 & 1\\
             \circled{1} &  100&  60&  25&  80& 226 & 0.953\\
             \circled{2}&  80&  55&  5&  60& 174 & 0.939\\
             \circled{3}&  80&  40&  5&  40& 144 & 0.921\\
             \circled{4}(a)&  70&  20&  70&  25& 144 & 0.909 \\
             \circled{4}(b)&  44&  44&  44&  44& 144 & 0.886 \\
             \circled{5}&  50&  25&  15&  25& 97 & 0.878 \\
             \bottomrule
        \end{tabular}
        \caption{Different combinations of downsample rate for body part and the consumed bandwidth}
        \label{tab:body-part-downsample-combinations}
    \end{minipage}\hfill
    \begin{minipage}{.45\textwidth}
        \centering
        \begin{tabular}{m{5.2cm}|C{2.1cm}}
            \toprule
            \textbf{Stage} & \textbf{Avg. Delay (ms)}\\ \midrule
            1. Capture a frame & 17.2 \\ \hline
            2. PC Construction & 6.3 \\ \hline
            3. Extract Skeleton & 67 \\ \hline
            4. Transmit to VABR server & 72\\ \hline
            5. Dynamic Calibration & 2.1 \\ \hline
            6. Skeleton Segmentation and Decimation & 167.5 \\ \hline
            \textbf{Overall} & \textbf{332} \\ \bottomrule
        \end{tabular}
        \caption{End-to-end introduced latency breakdown}
        \label{tab:end-to-end-latency}
    \end{minipage}
\end{table*}

\begin{figure*}[t]
    \begin{minipage}[t]{0.3\textwidth}
        \centering
        \includegraphics[width=\linewidth]{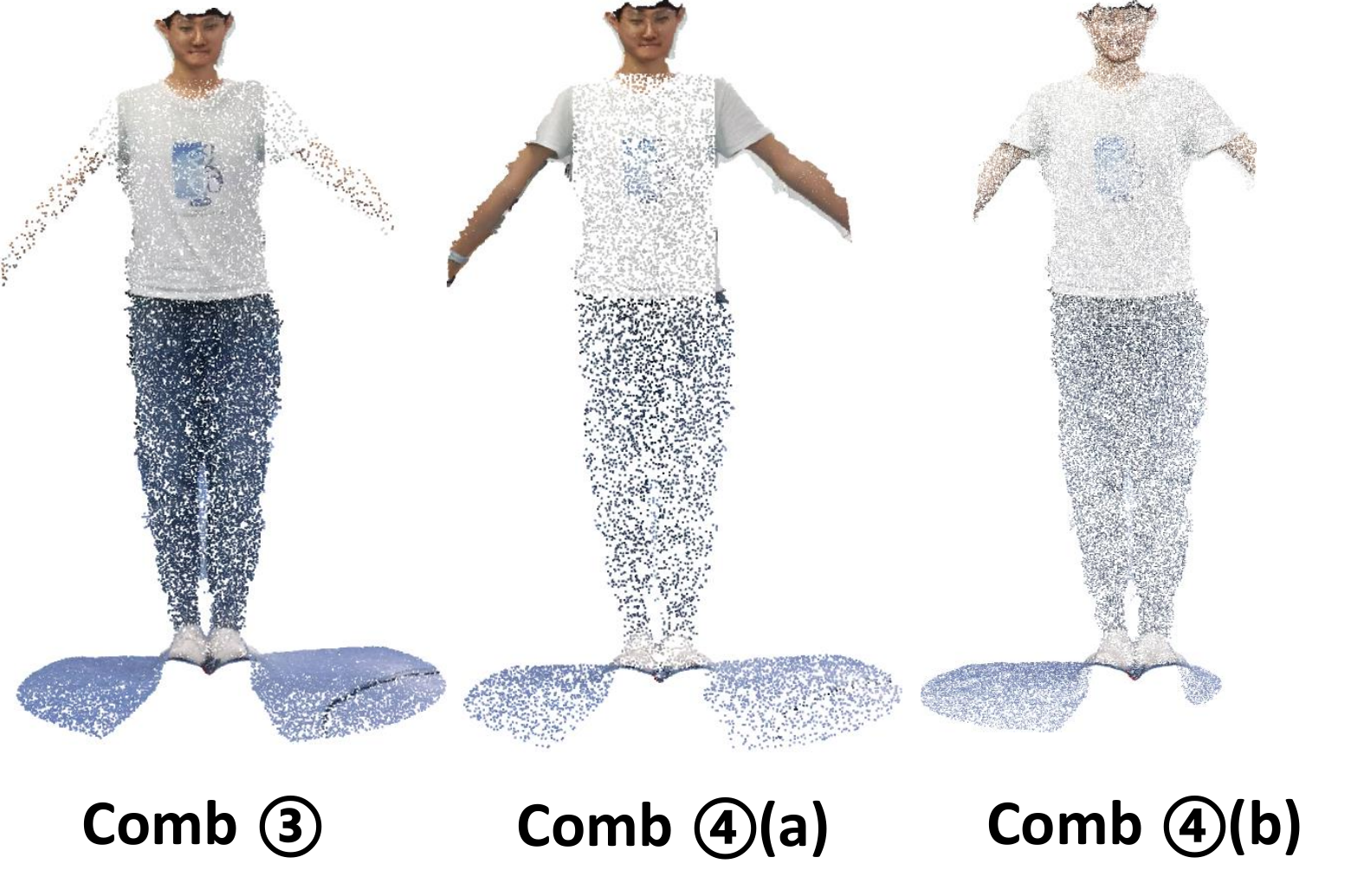}
        \caption{Visualized three combinations of downsample rates}
        \label{fig:combination-quality-compare}
    \end{minipage}
     \hfill
    \begin{minipage}[t]{0.34\textwidth}
        \centering
        \includegraphics[width=1\linewidth]{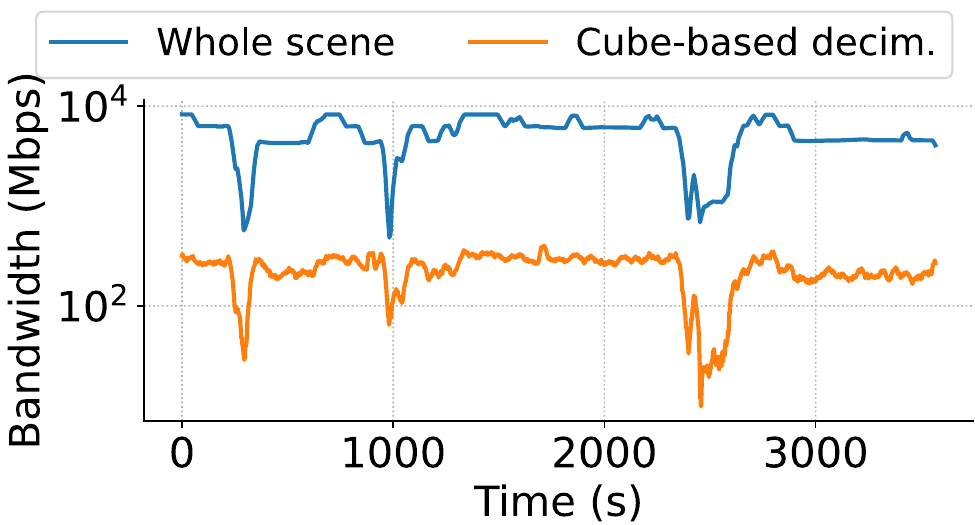}
        \caption{Bandwidth consumption of transmitting the whole and cube-based decimated scene}
        \label{fig:cube-scene-bandwidth-saving}
    \end{minipage}
     \hfill
    \begin{minipage}[t]{0.34\textwidth}
        \centering
        \includegraphics[width=1\linewidth]{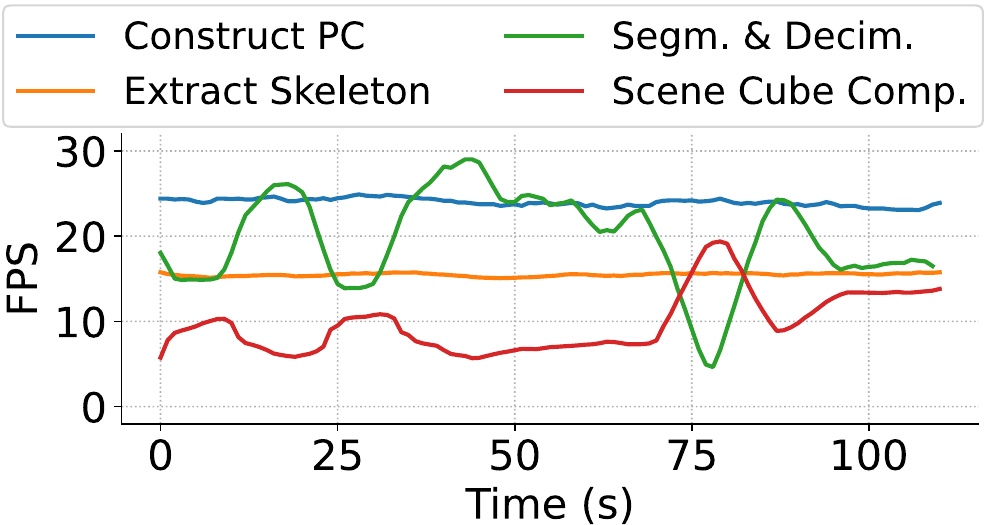}
        \caption{System throughput breakdown of compute-intensive steps}
        \label{fig:throughput-breakdown}
    \end{minipage}
\end{figure*}

We first show how skeleton-based segmentation and decimation affect the final visual quality of the whole body. We segment the body into four parts according to the integrity of the segmented part for ease of comparison. We carefully picked five combinations of the downsampling rates of each body part, while each of the combinations represents a different viewing concern: \circled{1}: clients with sufficient bandwidth; \circled{2}: keep head with high quality and downsample the others; \circled{3}\circled{4}: same bandwidth target, but downsample rate for each part selected reversely according to the quality decline trend (as given in Figure~\ref{fig:measure-visual-quality}); \circled{5}: clients with relatively low bandwidth.

We ordered the result by the consumed bandwidth of each combination, as shown in Table~\ref{tab:body-part-downsample-combinations}. As we decrease the amount of body data to be transmitted, the normalized SSIM gets worse. Comparing \circled{1} to \circled{5}, we found that transmitting those body parts with more points results in a globally worse visual quality. We also compare the rendered body of the combination \circled{3}, \circled{4}, and the original in Figure~\ref{fig:combination-quality-compare} to confirm that under the same bandwidth target, body parts with higher point count decrease the overall watching quality.

\subsection{Cube-based Scene Content Update}
We measure the bandwidth consumption of directly updating the whole volumetric scene and only updating the cube with high disparity. The resulting bandwidths of live streaming a two-minute volumetric video are illustrated in Figure~\ref{fig:cube-scene-bandwidth-saving}, from which we can tell a significant bandwidth saving of more than 22.5$\times$ on average.

\subsection{End-to-end Throughput and Latency}

We evaluate how our capture unit and VABR server could support four compute-intensive tasks of our proposed method. As shown in Figure~\ref{fig:throughput-breakdown}, the blue line indicates each capture unit stably constructs point clouds at an average of 24 FPS, while the body skeleton generates at 15.5 FPS.
Since skeleton extraction is the dependent task of skeleton-based segmentation, decimation, and scene composition, it restricted the throughput of the latter two tasks. We further recorded the 2-minute live capture, pre-extract the skeleton, and benchmarked them. Our results show that the performance of both tasks fluctuates from 4 to 29 FPS, as the person goes in, out, stays still, or moves around. %However, we found the reason behind the fluctuation of the two tasks is different: (1) for the skeleton-based segmentation and decimation task, only when the camera has captured a person will the server perform the task, (2) for the cube-based scene composition, a person entering into the scene blocks the camera's frustum to capture the content behind the person, leading to less points to process. %the skeleton-based segmentation and decimation can reach a stable throughput of 23.9 FPS while the scene comparison and composition float from 15 to 24 FPS. %We observe the reason for floating performance is the increase of processed points when the person moves in the captured scene

We next measure the time lag of our system. We found that processing frames from different capture units introduce different delays at the same timestamp, while the system output is determined by the slowest one. Table~\ref{tab:end-to-end-latency} gives the latency breakdown of each step $S_{1-6}$, and each value is calculated from the slowest processed frame across six cameras $C_{1-6}$, followed by averaging across total 3600 frames $F$, i.e. $AvgDelay_{S_i} = \sum_{f}^{F} \max(S_i(C_1,f), S_i(C_2,f), ..., S_i(C_6,f)) / F $. Our calibration only involves simple matrix computation on 32-point skeletons and uses an average of 2.1ms. The most time-consuming task is skeleton segmentation and decimation which contributes half of the latency.

\subsection{Ablation Study}

To validate the effectiveness of our proposed skeleton-based decimation mechanism, we conducted an ablation study. In this study, we meticulously examined the impact of replacing the skeleton-based decimation with a basic PDL down-sampling approach while maintaining the same network condition constraint. The experiment results are presented in Table~\ref{tab:body-part-downsample-combinations} and Figure~\ref{fig:combination-quality-compare}. In practice, for scenarios \circled{3}, \circled{4}(a), and \circled{4}(b), where the network bandwidth constraint was set to 144Mbps, we employed two different decimation mechanisms: skeleton-based for \circled{3}, \circled{4}(a), and basic downsampling for \circled{4}(b). Based on the results, we can tell that the adoption of basic downsampling resulted in a notable degradation in visual quality, as evident from the considerable drop in normalized SSIM. This finding underscores the inferiority of basic downsampling compared to the skeleton-based decimation approach. To enhance the comprehension of this disparity, Figure~\ref{fig:combination-quality-compare} offers a visual representation that illustrates how the skeleton-based decimation method excels in preserving the intricate details of the parts with higher saliency, even when subjected to identical network conditions.

%2-pages
%每个part分别evaluate

\section{Conclusion and Future Work}
In this paper, we present LiveVV, a holistic live volumetric video streaming system. LiveVV seamlessly integrates real-time capture, scene segmentation \& reuse, adaptive transmission, and immersive display to achieve end-to-end live streaming of volumetric video. To realize the system, we designed a body tracking-based calibration for instant deployment, selective data decimation \& reuse via static/dynamic segmentation, and a novel volumetric video adaptive bitrate streaming algorithm. Implemented on commodity hardware, LiveVV delivers volumetric video at 24 fps under 350ms latency, validating its ability to enable interactive live streaming applications. This pioneering system represents a major advance towards practical volumetric video streaming and lays the groundwork for future research in optimizing dynamic volumetric data capture, processing, and delivery.

Despite the significant progress in volumetric video streaming, there remain abundant opportunities for further research and development. The development of depth capture sensors and reconstruction algorithms can further enhance the capture quality and mobility, providing better visual quality. Additionally, optimizing compression techniques, particularly through the utilization of advanced neural compression methods can further enhance the steaming bandwidth.

% Lorem ipsum dolor sit amet, consetetur sadipscing elitr, sed diam
% nonumy eirmod tempor invidunt ut labore et dolore magna aliquyam erat,
% sed diam voluptua. At vero eos et accusam et justo duo dolores et ea
% rebum. Stet clita kasd gubergren, no sea takimata sanctus est Lorem
% ipsum dolor sit amet. Lorem ipsum dolor sit amet, consetetur
% sadipscing elitr, sed diam nonumy eirmod tempor invidunt ut labore et
% dolore magna aliquyam erat, sed diam voluptua. At vero eos et accusam
% et justo duo dolores et ea rebum. Stet clita kasd gubergren, no sea
% takimata sanctus est Lorem ipsum dolor sit amet. Lorem ipsum dolor sit
% amet, consetetur sadipscing elitr, sed diam nonumy eirmod tempor
% invidunt ut labore et dolore magna aliquyam erat, sed diam
% voluptua. At vero eos et accusam et justo duo dolores et ea
% rebum.

%% if specified like this the section will be committed in review mode
\acknowledgments{
The authors wish to thank A, B, and C. This work was supported in part by
a grant from XYZ.}

\bibliographystyle{abbrv-doi}

\bibliography{template}
\end{document}